%% file: manuscript.tex
\documentclass[sigconf]{acmart}

\usepackage{graphicx}
\usepackage{array}
\usepackage{color}

\newcommand{\added}[1]{{#1}}
\newcommand{\deleted}[1]{}

\newcommand{\hazel}[1]{{#1}}
\newcolumntype{C}[1]{>{\centering\let\newline\\\arraybackslash\hspace{0pt}}m{#1}}
\newcolumntype{P}[1]{>{\centering\arraybackslash}p{#1}}
\AtBeginDocument{%
  }

\copyrightyear{2025}
\acmYear{2025}
\setcopyright{cc}
\setcctype{by}
\acmConference[CHI '25]{CHI Conference on Human Factors in Computing Systems}{April 26-May 1, 2025}{Yokohama, Japan}
\acmBooktitle{CHI Conference on Human Factors in Computing Systems (CHI '25), April 26-May 1, 2025, Yokohama, Japan}\acmDOI{10.1145/3706598.3713430}
\acmISBN{979-8-4007-1394-1/25/04}
\begin{document}

\title[Goal Exploration for Gender-Affirming Voice Training]{Beyond the ``Industry Standard'': Focusing Gender-Affirming Voice Training Technologies on Individualized Goal Exploration}

\author{Kassie Povinelli}
\affiliation{%
  \institution{University of Wisconsin-Madison}
  \city{Madison}
  \state{Wisconsin}
  \country{USA}
}
\email{kassie.povinelli@wisc.edu}
\orcid{0009-0006-6969-1622}
\author{Hanxiu ``Hazel'' Zhu}
\affiliation{%
  \institution{University of Wisconsin-Madison}
  \city{Madison}
  \state{Wisconsin}
  \country{USA}
}
\email{hzhu339@wisc.edu}
\orcid{0000-0003-0777-6090}
\author{Yuhang Zhao}
\affiliation{%
  \institution{University of Wisconsin-Madison}
  \city{Madison}
  \state{Wisconsin}
  \country{USA}
}
\email{yuhang.zhao@cs.wisc.edu}
\orcid{0000-0003-3686-695X}

\renewcommand{\shortauthors}{Povinelli et al.}

\begin{abstract}
Gender-affirming voice training is critical for the transition process for many transgender individuals, enabling their voice to align with their gender identity. Individualized voice goals guide and motivate the voice training journey, but existing voice training technologies fail to define clear goals. We interviewed six voice experts and ten transgender individuals with voice training experience (voice trainees), focusing on how they defined, triangulated, and used voice goals. We found that goal voice exploration involves navigation between \deleted{approximate}\added{descriptive} and \deleted{clear}\added{technical} goals, and continuous reevaluation throughout the voice training journey. Our study reveals how \added{goal descriptions, subjective satisfaction,} voice examples\deleted{, character descriptions}, and voice modification and training technologies inform goal exploration, and identifies risks of overemphasizing goals. We identified technological implications informed by \deleted{the }\added{existing expert and trainee strategies,}\deleted{separation of voice goals and targets,} and provide guidelines for supporting individualized goals throughout the voice training journey based on brainstorming with trainees and experts.
\end{abstract}




\maketitle

\section{Introduction}
\input{sections/introduction}
\section{Related Works}
\input{sections/related_works}
\section{Method}
\input{sections/methods}
\section{Findings}
\input{sections/findings_revised}
\section{Discussion}

\input{sections/discussion}
\section{Conclusion}

\input{sections/conclusion}
\begin{acks}
\input{sections/acknowledgements}
\end{acks}
\bibliographystyle{ACM-Reference-Format}
\bibliography{Related_Works_Bibliography/Current_Technologies,Related_Works_Bibliography/thematic_analysis,Related_Works_Bibliography/transitioning_online,Related_Works_Bibliography/voice_changer_research,Related_Works_Bibliography/voice_changer_technologies,Related_Works_Bibliography/Voice_Training_Background,Related_Works_Bibliography/Voice_Training_New_Technology_Research,Related_Works_Bibliography/VR_research,Related_Works_Bibliography/trans_technologies}
\appendix
\newpage
\input{sections/Appendix}
\end{document}

%% file: sections/introduction.tex
\added{
Voice plays a crucial role in gender expression and identity affirmation for transgender and gender-diverse (TGGD) individuals \cite{Dahl2020, Buckley2020}. However, many TGGD individuals face significant challenges in aligning their voices with their gender identity \cite{trans_women_effects_of_testosterone_on_voice, transmasc_hormone_therapy_voice, my_paper}. 
To address this challenge, many TGGD people pursue gender-affirming \textit{voice training}---a comprehensive process of developing vocal muscle control, modifying speech patterns, and acquiring new voice qualities (e.g., pitch, resonance, weight) to better align their voices with their identities \cite{Gelfer2013_exercise, Stemple2000-ku, Gelfer2013_perceptual, Carew2007, voice_perceptions_QOL}.}
\added{
Voice training traditionally involves collaboration with voice experts (e.g., speech-language pathologists (SLPs), voice coaches) who provide personalized guidance and feedback to help trainees achieve desired voice \cite{giving_voice_to_the_person_inside, interdisciplinary_voice_workshop}. Various technologies have been used by both experts and trainees to support voice training, such as spectrograms and pitch analyzers during sessions, and client-side applications to support teletherapy and independent practice (e.g., smartphone-based voice analysis apps) \cite{teletherapy_voice_training, teletherapy_for_voice}.
}

\added{
Identifying voice goal is a critical early step of voice training, where the TGGD trainees and experts work together to develop a desired voice to guide the entire voice training journey \cite{Chadwick2022, connected_speech_pathology, john_hopkins_medicine}. 
However, this process is challenging since each TGGD individual may have unique identity representation needs based on their demographics and desires \cite{giving_voice_to_the_person_inside}. 
Despite growing recognition of diverse and individualized needs for voice goals \cite{trans_competent_interaction_design, haimson_how_2023}, 
current voice training technologies predominantly enforce standardized, binary gender-based voice goals \cite{voice_training_software_considerations, voice_training_app}. For example, applications like Voice Pitch Analyzer \cite{voicepitchanalyzer2024} and Voice Tools \cite{voice_tools} measure the user's voice against predefined feminine and masculine pitch ranges. 
This rigid approach poses significant barriers for TGGD individuals whose ideal voices may not align with prescribed norms. 
There is a critical gap between the diverse voice needs of TGGD users and the arbitrary binary-gender-based technologies, underscoring the needs for more TGGD-centered voice training technologies to enable authentic, individualized voice goal exploration. 
}

\added{To inspire technology design, we seek to deeply understand the landscape of personalized voice goals from both TGGD individuals' and experts' perspectives. Specifically, we focus on three key research questions:
\begin{itemize}
    \item[\textbf{RQ1.}] How do TGGD individuals identify and develop their voice goals and what challenges do they face?
    \item[\textbf{RQ2.}] How do voice goals impact the broader voice training journey for both trainees and experts? 
    \item[\textbf{RQ3.}] What are the technological design implications for supporting personalized goal exploration and voice training?
\end{itemize}
}
\added{
To answer these research questions, we conducted interview studies with six \textit{voice experts} and 10 \textit{TGGD voice trainees}, 
focusing on voice training strategies, evolving understandings of voice goals, and technology use throughout the voice training journey. 
Through technology brainstorming, we put additional focus on voice changers based on their potential for identity exploration and voice training support \cite{my_paper} and recent advances in signal-processing-based \cite{permod_robin} and AI-based \cite{RVC_WebUI, w-okada_voice_changer} real-time voice changing technologies, as well as social VR based on its potential as a space for practicing voice in new social contexts \cite{freeman_rediscovering_2022, my_paper}, using them as technology probes to promote idea generation. 
}

Our research revealed key factors that define voice goals, challenges in voice goal identification, and the role of voice goals in driving the voice journey. 
\added{First, we discovered a significant communication gap between trainees and experts:
trainees envision their ideal voices via \textit{layman-descriptive goals} (e.g., character descriptions, desired vocal utility, comfort with the voice), while experts rely on actionable \textit{technical goals} with voice characteristics (e.g., pitch, resonance) to design training plans. As a result, subjective satisfaction and voice examples served as a crucial bridge between these different ways of conceptualizing voice in voice goal exploration. Second, we found that goals naturally evolve from approximate motivations to clear technical trajectories during voice training, alongside trainees' increasing understandings of voice and their identities, which requires technology support to adapt with this evolution. Third, we identified that while clear technical goals provide essential direction, rigidity around and over-emphasis of end-goals in voice training demotivates and stalls voice training progress.}


\added{Based on these insights, we develop design guidelines for technologies that support voice goal exploration and the overall voice training journey. Through these contributions, our research sets a foundation for voice training technologies that center individual identity and authentic self-expression rather than standardized binary goals. 
}

%% file: sections/related_works.tex
\subsection{\added{Need-Driven Technology Development for the TGGD Community}}
\added{
Technology development targeting transgender and gender diverse (TGGD) users has often fallen short of meeting community needs \cite{voice_training_app, bivens_gender_2017, haimson_how_2023}, despite its critical role in identity exploration, healthcare access, and daily life. Recent research reveals significant gaps between design intentions and community requirements, highlighting the importance of meaningful TGGD involvement throughout the technology development process \cite{chuanromanee_transgender_2021, open-source_voice_training_app}. In this section, we introduce prior technology design and practices for the TGGD community in HCI literature. 
}
\subsubsection{\added{Inclusive Design Practices}}
\added{
The development of trans technologies requires meaningful community involvement throughout the design process, yet current approaches often fall short of this standard. Prior research has highlighted this issue by investigating the barriers and needs of transgender users and the technology design process for them \cite{open-source_voice_training_app, haimson_how_2023, chuanromanee_transgender_2021, bivens_gender_2017, devito_how_2022}. 
For example, Ahmed et al. \cite{open-source_voice_training_app} introduced the concept of trans-competent interaction design through a qualitative study, examining voice, identity, and technology use among trans individuals. Through this framework, they demonstrated how technology design must challenge traditional assumptions about gender while creating systems that actively support transgender users' needs and help reduce the social barriers they face. 
Building on these principles, Haimson et al. \cite{haimson_how_2023} conducted a comprehensive analysis of trans technology design processes through 104 interviews with 115 creators. 
They identified a significant gap between design and deployment, highlighting how even well-designed technologies often fail to reach the communities they are intended to serve.}
\added{Additionaly, through interviews with 21 transgender individuals, Chuanromanee and Metoyer \cite{chuanromanee_transgender_2021} conducted an in-depth investigation of trans people's technology needs related to health and transition, 
highlighting how current technologies often fail to account for the non-linear nature of transition and the varied ways individuals navigate their gender identity.
%
}
\subsubsection{\added{Supporting Identity Presentation and Expression}}
\added{
Identity representation is a critical need for TGGD individuals, who increasingly turn to virtual environments not only to explore and express their authentic selves, but also to build confidence and develop strategies for identity presentation in their everyday lives \cite{doyle_transgender_2022, cavalcante_i_2017, buss_transgender_2022, haimson_disclosure_2015, bessiere_ideal_2007, pickering_second_2018}.
Through ethnographic research, Cavalcante \cite{cavalcante_i_2017} illustrated how transgender individuals increasingly turn to online spaces for identity exploration and management, finding these digital environments offer unique opportunities for controlled identity presentation while navigating everyday life. 
In examining social media platforms, Buss et al. \cite{buss_transgender_2022} found that transgender individuals actively curate their online presence across multiple platforms and accounts, strategically managing different aspects of their identity presentation for varied audiences. 
Beyond social media, Bessière et al. \cite{bessiere_ideal_2007} and Pickering \cite{pickering_second_2018} showed how game environments, like World of Warcraft and Second Life, enable users to explore idealized versions of themselves through character creation, leading to authentic identity development.
}
\added{The emergence of social VR has introduced new opportunities and challenges for identity expression. 
Freeman et al. \cite{freeman_rediscovering_2022} specifically examined how non-cisgender individuals navigate these novel spaces, revealing how social VR enables new forms of gender identity expression through embodied interactions. 
Povinelli and Zhao \cite{my_paper} built on this work, investigating how TGGD users achieve a comprehensive identity representation by combining voice modification through voice changer applications alongside avatar-based identity representation. They found that the ``passing'' experiences in social VR not only reduce harassment and increase well-being, but also serve as motivations for taking steps to transition outside of the virtual world, such as accessing gender-affirming healthcare and voice training. 
}
\subsection{Gender-Affirming Voice Training}
Gender-Affirming voice training is a process through which TGGD individuals (i.e., voice trainees) alter their voices to better match with their identities. This process typically involves altering characteristics of voice, such as pitch, resonance, weight, and fry to best match an individual's voice to their identity through exercise targeting physiological characteristics, such as the vocal muscles \cite{Gelfer2013_exercise, Stemple2000-ku}, and psychological characteristics, such as habits around using one's voice \cite{Gelfer2013_perceptual, Carew2007}. Through voice training, TGGD individuals can overcome \textit{voice dysphoria}, gender dysphoria caused by incongruence between one's voice and identity \cite{prismaticspeech_dysphoria}, while also increasing safety and reducing social stigma \cite{Oates2023}. For transfeminine individuals, this involves feminizing characteristics of their voices, such as pitch, resonance, and vocal weight to undo the masculinizing effects of testosterone \cite{Dahl2020}. For transmasculine individuals, this process involves masculinizing similar characteristics of voice \cite{Buckley2020}. While testosterone-based hormone therapy effectively masculinizes characteristics of voice such as pitch and resonance \cite{HodgesSimeon2021}, transmasculine individuals may not want to pursue hormone therapy \cite{Puckett2017}, and the masculinizing effects of testosterone alone may not be sufficient for aligning voice to identity \cite{Papeleu2024}.
\subsubsection{Voice Training Expert Methodology}
Gender-affirming voice training involves building knowledge and identifying vocal goals \cite{Merrick2022}, altering voice physiology through the consistent use of exercises \cite{Gelfer2013_exercise}, and building vocal habits to generalize the ideal voice outside of practice sessions \cite{giving_voice_to_the_person_inside}.
Many voice trainees work with a voice expert, such as a speech-language pathologist (SLP) or voice coach.
Voice experts play a critical role in identifying individualized and achievable goals based on measures of satisfaction of voice and features of the current voice \cite{giving_voice_to_the_person_inside}, building trainee knowledge and self-competence with voice training \cite{Behlau2023}, helping the trainee maintain vocal health and avoid strain \cite{Behlau2023}, and counseling the trainee through generalizing the voice-in-training  beyond the therapy session \cite{Leyns2022}. Although part of this can be done in session, the limited time of sessions requires that experts prepare trainees to effectively practice voice outside of sessions through directed voice exercises \cite{Iwarsson2014}, regular practice to build habits \cite{Behlau2023, Iwarsson2014}, and building self-perception of voice \cite{Iwarsson2014}.
\subsubsection{Voice Training in Community}
For many trainees, the financial cost of working with a voice expert presents a barrier to voice training. To overcome this barrier, many trans people join gender-affirming voice training communities, including in-person workshops \cite{interdisciplinary_voice_workshop} or online social media groups, such as those on Discord \cite{discord_transvoice_2021}, Reddit \cite{reddit_transvoice}, or VRChat \cite{trans_academy_voice_training_group}. These communities  provide knowledge through databases of voice training resources \cite{sumian_voice_wiki, reddit_transvoice}, hone skills and provide feedback through group voice training sessions and crowdsourced evaluations of voice training clips, and provide emotional support \cite{safe_space_lgbtq_social_vr}.
\subsection{Voice Training Technology Developments and Trans-Inclusive Technology Design}
While voice training technologies can accelerate progress in voice training, they lack the ability to measure important vocal characteristics and provide limited guidance for target-setting \cite{voice_training_software_considerations}. \added{Recent voice training technology research focuses on the limitations of technologies for guiding individualized voice training journeys, identifying how applications rely on rigid pre-defined gender-binary trajectories, and exploring ways to accommodate unique trajectories and goals through customizability beyond standard gender-based criteria \cite{voice_training_app, open-source_voice_training_app, voice_training_software_considerations}.}

\subsubsection{Technology Use by Experts}
To aid in evaluation of voice and setting precise targets, experts use voice analysis tools in-session, including voice analysis software like Praat \cite{praat} and Visi-pitch \cite{visi-pitch}, tone generators, and spectrograms. To expand access to gender-affirming voice therapy, experts may use telepractice, which produces similar outcomes compared to in-person sessions \cite{teletherapy_for_voice, teletherapy_voice_training}. To aid in out-of-session practice, experts guide trainees on how to use voice analysis tools such as smartphone-based pitch analyzers and practice applications such as Voice Tools \cite{voice_tools}.

\subsubsection{\added{Individualized Goals in Voice Technology Design}}
New voice training technology design research focuses on \added{addressing the limitations of existing voice training applications, which set targets and measure progress against idealized voice standards through tools that can accommodate the wide spectrum of goals and experiences within the TGGD community \cite{trans_competent_interaction_design}}. Ahmed et al. conducted a comprehensive analysis of \added{five }mobile voice training applications, revealing that they reinforce binary gender norms and societal expectations \added{of passing as cisgender, struggle to accommodate the cultural needs of users, and use language promoting individualized journeys while accomodating only rigid pre-defined goal ranges} \cite{voice_training_app}. In response to these limitations, they  explored the potential of community-driven, free and open-source software development for gender-affirming voice training\added{, placing customizability as a defining guideline alongside vocal health and affirmation} \cite{open-source_voice_training_app}. Building on this, Bush et al. analyzed TGGD individuals' experiences with voice training software, and \added{identified automated personalized target-setting as a key feature alongside feedback, accountability mechanisms, and training for voice characteristics beyond pitch, showing that while TGGD users desired high customizability for voice training trajectories, they also wanted software which could automate target-setting along individualized trajectories, performing the role of a voice coach or therapist} \cite{voice_training_software_considerations}. 
\input{Tables/expert_demographics}
\subsubsection{\added{Voice Changing as a New Foundation for Voice Technology Design.}}
Further improvements in \added{signal-based and AI} voice \added{changer} technologies enable more in-depth analysis and exploration of TGGD voices\added{, providing a testbench for real-time exploration of voice goals accessible to TGGD individuals without expert guidance.}
Enhancement-based \cite{MorphVOX_Pro, SuperVoiceChanger, Clownfish} and AI-based \cite{Voicemod, RVC_WebUI, w-okada_voice_changer} voice changers allow TGGD users to present their gender identities online \added{by shifting their voice characteristics to match a desired voice presentation}. When combined with virtual representations of one's identity, such as personalized avatars in social virtual reality applications \cite{freeman_body_2021, freeman_rediscovering_2022}, TGGD individuals can internalize the modified voice through experiences with others, allowing the modified voice to become both a motivation for voice training and voice goal \cite{my_paper}. While AI voice changer technologies, such as retrieval-based voice conversion (RVC) voice changers, reliably produce natural-sounding voices, the ``black-box'' nature of these technologies prevents users from understanding the underlying voice characteristics of generated voices. Netzorg et al. develop a perceptually-based voice modification tool with inputs mirroring the ways that speech-language pathologists describe voice, enabling voice generation grounded in identifiable voice qualities \cite{permod_robin}. Goal voice exploration leveraging voice changer technologies \cite{my_paper} and \added{perceptually grounded voice measurement \cite{acousticgenderspace} and modification \cite{permod_robin}} could provide authentic and clear trajectories for voice training \added{accessible directly to trainees.}

While goal voice exploration is an important initial step in the voice training journeys of TGGD people, \added{existing voice training technologies lack the individualized goal exploration and guidance central to expert-led voice therapy.}  Given the importance of individualized and authentic goals to inclusive voice training technology development \cite{open-source_voice_training_app, voice_training_software_considerations, trans_competent_interaction_design}, there is a need for  technology design research addressing \added{how voice training technologies can incorporate successful expert and trainee strategies for developing personalized voice training journeys.} 

%% file: Tables/expert_demographics.tex
\begin{table*}[h]
\renewcommand{\arraystretch}{1.3}
\centering
\caption{Voice Expert Demographics}
\Description[Demographics of the six experts included in the study]{This table presents demographic information for six voice experts (E1-E6). The table contains seven columns: ID, transgender status, gender-nonconforming status, gender identity, age, years of experience, and notes. The experts range in age from 23 to 47 years old, with experience levels ranging from 2.5 to 18 years. Two experts identify as transgender and gender-nonconforming, both being nonbinary. Three experts identify as female, one as male, and two as nonbinary. One expert (E4) has notable dual-language experience. The data reveals diversity in both demographic characteristics and professional experience among the voice experts.}
\label{tab:expert_demographics}
\begin{tabular}{C{0.05\linewidth} C{0.1\linewidth} C{0.1\linewidth} C{0.15\linewidth} C{0.1\linewidth} C{0.1\linewidth} C{0.2\linewidth}}
\toprule
ID & Transgender? & Gender-Nonconforming? & Gender Identity & Age & Years of Experience & Notes \\
\midrule
E1 & No & No & Female & 47 & 18 & N/A\\
\hline
E2 & Yes & Yes & Nonbinary & 30 & 3 & N/A\\
\hline
E3 & No & No & Female & 33 & 9 & N/A\\
\hline
E4 & No & No & Female & 23 & 3 & Dual-Language Experience\\
\hline
E5 & No & No & Male & 27 & 2.5 & N/A\\
\hline
E6 & Yes & Yes & Nonbinary & 35 & 8& N/A\\
\bottomrule
\end{tabular}
\end{table*}

%% file: sections/methods.tex

To better understand how voice goals, voice practice methods, and voice generalization strategies interact with technology use, we conducted an interview study involving gender-affirming voice experts and \added{TGGD} trainees with at least three months of voice training experience.

\subsection{Participants}
We recruited two groups of participants, six gender-affirming voice experts, including five speech-language pathologists (SLPs) and one SLP-in-training, and ten TGGD voice trainees, comprising transgender individuals at various stages of their voice training journeys.



\subsubsection{Voice Experts}
We recruited six voice experts, with ages ranging from 27 to 47 (\textit{M} = 32.5, \textit{SD} = 7.56) and years of experience ranging from 2.5 to 18 (\textit{M} = 7.25, \textit{SD} = 5.44). Among the voice experts, two (E2, E6) identified as transgender and/or nonbinary, one (E5) was a cisgender man, and three (E1, E3, E4) were cisgender women. Table \ref{tab:expert_demographics} shows voice experts' demographic information.
We emailed expert recruitment information to gender-affirming voice therapy programs at various universities and through our own university's research email recruitment system. To be eligible for participation, voice experts had to be at least 18 years old and have experience conducting gender-affirming voice therapy or coaching with transgender clients.

\subsubsection{Voice Trainees}
We recruited 10 TGGD voice trainee participants  with ages ranging from 19 to 68 (\textit{M} = 31.3, \textit{SD} = 13.88). The trainees represented a diverse range of gender identities and voice training goals, including masculinization, feminization, and androgenization. The group comprised one (T1) nonbinary transfeminine participant, one (T6) nonbinary transmasculine participant, six trans women (T3-T5, T7-T9), and two trans men (T2, T10). T2 and T10 noted that testosterone-based hormone therapy had lowered their voices, while T6 mentioned that they were not on testosterone. T3 mentioned that she started puberty blockers before adulthood, but already had her voice drop because of testosterone-based puberty. Both T2 and T10 are bilingual, with experiences that reflect voice training in two languages. Participants had varying levels of voice training experience, with some independently working on voice training and others working with voice experts such as voice coaches or SLPs. Table \ref{tab:trainee_demographics} shows voice trainees' demographic information.

We emailed trainee recruitment information through our university's research email recruitment system and created recruitment postings on multiple transgender identity-focused and voice training-focused Discord servers. Participants were eligible if they were at least 18 years old and had at least three months of consistent gender-affirming voice training experience, either alongside an expert or without an expert. 

\input{Tables/trainee_demographics}

\subsection{Procedure}
We conducted 1-1.5 hour single-session \added{semi-structured} interviews with gender-affirming voice experts via zoom. We first gathered information about age, gender identity, professional role, and years of experience. Then, \added{we asked about their strategies, challenges, and technology use in different voice training stages, with more emphasis on the goal exploration phase.} Specifically, we started by asking experts to describe their initial goal voice understanding and exploration process with their clients as well as strategies and technologies used in this process. We also  asked about the methods for measuring voice suitability and specific aspects of voice typically trained. \added{We then moved to the overall voice training process}, discussing common pain-points in voice training efforts and the use of voice-training-specific technologies (e.g., pitch analyzers, spectrograms) and more general technologies benefiting voice training (e.g., voice-chat and social VR applications) inside and outside sessions. We finally focused on how voice therapists aided clients in generalizing their voice use, transitioning from exercises to using their voice-in-training in social environments. 

\added{We ended the interview with a semi-structured brainstorming section focused on the potential integration of emerging voice technologies in gender-affirming voice training. Using state-of-the-art voice changer applications \cite{w-okada_voice_changer, RVC_WebUI}, social VR platforms (previously explored in \cite{my_paper, VR_for_voice_training_and_conversation_simulation, freeman_rediscovering_2022}), and AI-assisted training tools as discussion probes, participants discussed how these technologies could be leveraged to support the goal exploration process and general voice training journey. To better prompt the participants, we asked about their technology ideas for different voice training phases, from voice example exploration, to intermediate training milestones, to generalizing voice to real life. We also consistently referred back to their prior technology experiences to inspire them and facilitate their ideation.}

We conducted 1-2 hour single-session \added{semi-structured} interviews with TGGD trainees via Zoom. We first gathered information about age, gender identity, voice training experience, and work with voice experts. \added{We then focused on understanding the trainee's voice goal exploration process}. Specifically, we asked how trainees identified and modified their voice goals, what factors informed these goals, how voice goals impacted and shaped overall voice training journey, and how social factors and technology use affect their voice goal identification.

\added{Mirroring our expert interviews, we finally conducted a semi-structured brainstorming section to discuss potential integration of emerging voice technologies in their training journey, 
seeking to understand how these technologies might complement and enhance their voice training experiences.}

\subsection{Data Recording and Analysis}
All interviews were conducted and recorded by a single researcher using Zoom. We used Zoom's automatic transcription service for transcription. To ensure accuracy, we listened to the audio recordings while coding the transcripts, making corrections to the transcripts as necessary. Throughout the interview process, we made iterative improvements to the brainstorming section of our interviews. We analyzed transcripts using thematic analysis \cite{thematic_analysis}.

We selected three representative transcripts from trainees and two representative samples from experts, \added{both of }which two researchers open-coded independently \added{using in-vivo codes}. Then, the two researchers worked together to agree on codes for the open-coded transcripts\added{, discuss emerging themes,} and \added{develop} two codebooks, one for experts and one for trainees. They then used these refined codebooks alongside continued independent open-coding for the remaining transcripts. To maintain coding congruence, the two researchers regularly checked each other's codes and open-codes and updated the codebooks for trainees and experts.

\added{Using an affinity diagram \cite{Lisle2020, kawakita1991original}, we clustered and categorized the codes based on emerging patterns using both deductive analysis from our research questions and inductive analysis alongside memos to identify unexpected patterns. To discover subthemes and themes that crossed between the categories and sub-categories in the diagram, we collaboratively identified codes and categories that regularly occurred together in the transcripts and identified relationships between categories upon agreement between the two researchers. 
With this method, we first developed initial subthemes and themes for the trainees and experts separately. Then, we compared the themes between trainees and experts, combined common themes, and finalized the themes and subthemes across participants.
Finally, we validated and refined the themes by cross-referencing with the original codes, transcripts, and memos. As a result, we identified eight shared themes relevant to voice goals and four themes relevant to designing goal exploration technologies, alongside several emerging themes for continued research.}

\input{Tables/expert_technology_use}
\subsection{Positionality Statement}
We acknowledge how the intersection of our identities with our roles as researchers influence our work. The primary researcher, a transgender woman, drew on her lived experience to inform the study design, interview approach, and data interpretation. The second researcher, a cisgender woman with an accessibility research background, contributed to coding, analysis, and background research. The third researcher, a cisgender woman with expertise in accessibility research, ensured research quality and rigor.

%% file: Tables/trainee_demographics.tex
\begin{table*}[h]
\renewcommand{\arraystretch}{1.3}
\caption{Voice Trainee Demographics} 
\Description[Demographics of the ten trainees included in the study]{This table presents demographic data for ten voice trainees (T1-T10). The table tracks eight characteristics: ID, age, gender identity, gender-nonconforming status, voice training experience, progress estimate, whether they trained with an expert, and additional notes. Trainees range in age from 19 to 68 years old, with varying gender identities (including female, male, nonbinary transfeminine, and nonbinary transmasc). Training experience spans from 8 months to 3-4 years, with progress levels from "inconsistent" to "finished." Four trainees identify as gender-nonconforming. Notable additional characteristics include ADHD (3 trainees), autism, hard of hearing, and specific ethnic identifications (Hispanic, Hmong).}
\label{tab:trainee_demographics}
\begin{tabular}{C{0.05\linewidth} C{0.1\linewidth} C{0.15\linewidth} C{0.12\linewidth} C{0.1\linewidth} C{0.1\linewidth} C{0.1\linewidth} C{0.1\linewidth}}
\toprule
ID & Age & Gender Identity & \added{Gender-Nonconforming?} & Voice Training Experience & Progress Estimate & Trained with Expert? & \added{Notes} \\
\midrule
T1 & 36 & Nonbinary transfeminine & \added{Yes} & 2 years & Inconsistent & Yes & \added{N/A}\\
\hline
T2 & 24 & Male  & \added{Yes} & 2-3 years & 75\% through & No & \added{Hispanic, ADHD}\\
\hline
T3 & 20 & Female & \added{No} &  3-4 years & Near-done & No & \added{N/A}\\
\hline
T4 & 19 & Female & \added{No} &  3 years &Near-done & No & \added{ADHD}\\
\hline
T5 & 25 & Female & \added{No} & 1-2 years & Midway & No & \added{N/A}\\
\hline
T6 & 20 & Nonbinary transmasc & \added{Yes} & 2 years & Ongoing & No & \added{ADHD}\\
\hline
T7 & 38 & Female &\added{No} & 8 months &  Near midway & Yes & \added{N/A}\\
\hline
T8 & 68 & Female & \added{No} & 2 years &  Ongoing & Yes & \added{Hard of hearing}\\
\hline
T9 & 29 & Female & \added{No} &  2 years &Finished & No & \added{Autistic}\\
\hline
T10 & 34 & Male & \added{No} & 2 years & Midway & No & \added{Hmong}\\
\bottomrule
\end{tabular}
\end{table*}

%% file: Tables/expert_technology_use.tex
\begin{table*}[h]
\renewcommand{\arraystretch}{1.3}
\centering
\caption{Technology Use for Voice Therapy Support Among Expert Participants}
\Description[Table of technologies used by experts]{This table details the technology tools used by six voice experts (E1-E6) across four categories: assessment tools, practice tools, self-monitoring tools, and generalization tools. Assessment tools commonly include spectrograms and voice analysis software like Praat. Practice tools range from basic piano apps to specialized voice analysis software. Self-monitoring tools typically include voice recording capabilities and various reminder systems (calendars, alarms, visual reminders). Generalization tools vary significantly, with some experts using virtual reality platforms (VRChat), while others utilize AI-based tools or don't report any specific tools. The data shows a wide range of technological approaches to voice therapy support, from basic recording tools to advanced virtual environments.}
\label{tab:voice-tech-usage}
\begin{tabular}{C{0.05\linewidth} C{0.2\linewidth} C{0.2\linewidth} C{0.2\linewidth} C{0.2\linewidth}}
\toprule
ID & Assessment Tools & Practice Tools & Self-Monitoring Tools & Generalization Tools \\
\midrule
E1 & Praat, Spectrogram & Voice Analyzer & Online journal, Health tracking, Calendar, Voice recording & VRChat, Social VR, Sumian Voice \\
\hline
E2 & Voice analysis apps, Spectrogram & Piano app, Voice Tools& Voice recording, Digital reminders, Calendar, Visual reminders (post-its)& Large-language-model (LLM)-generated content \\
\hline
E3 & Digital analyzer, Frequency analyzer & Audio recording tools & Visual reminders (post-its) & N/A \\
\hline
E4 & Visi-pitch, Pitch monitor & Piano app, Voice mixer & Pitch grabber & Teletherapy tools \\
\hline
E5 & Computerized Speech Lab (CSL), Spectrogram & Voice Tools & Voice recording, Real-time feedback & N/A \\
\hline
E6 & Praat & Voice Analyst, Voice Tools& Voice recording, Voice Memo, Alarms/Reminders, Calendar, Visual reminders (post-its) & VTuber, Character AI, Zoom \\
\bottomrule
\end{tabular}
\begin{flushleft}
\end{flushleft}
\end{table*}

%% file: sections/findings_revised.tex
\input{Tables/trainee_technology_use}

\begin{figure*}[htbp]
    \centering
    \includegraphics[page=4, width=\textwidth, trim={0 4.3cm 0 3cm},clip]{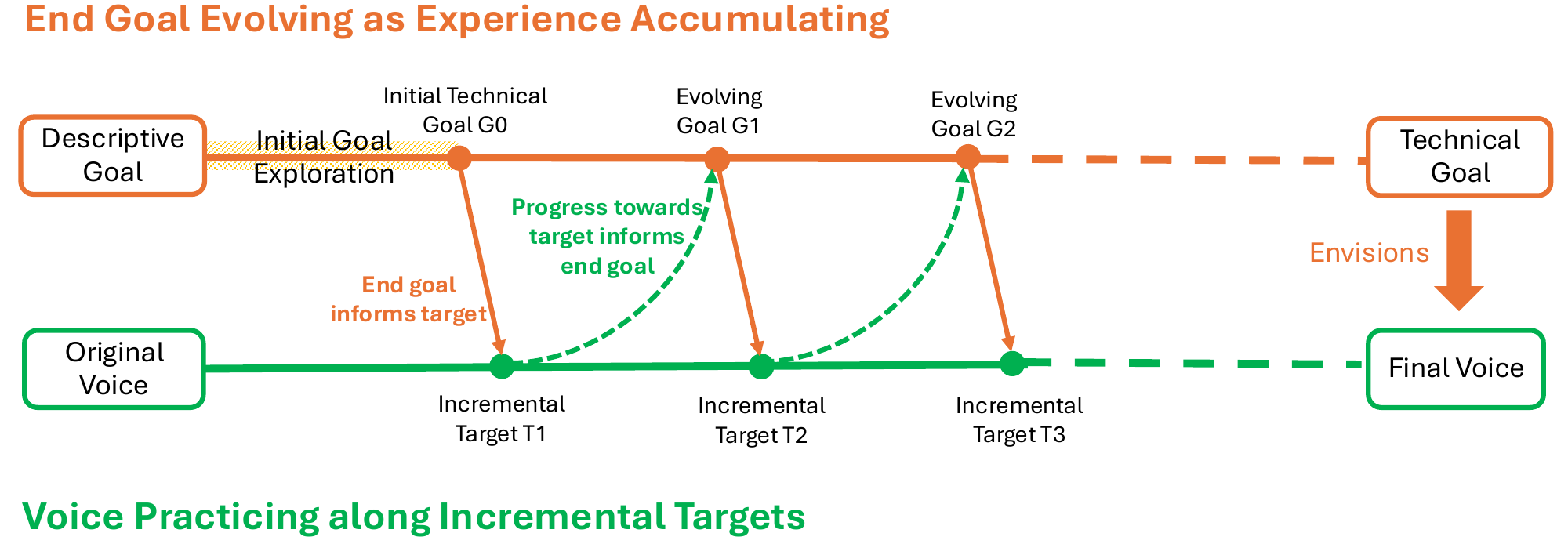}
    \Description[Goal Evolution Timeline Visualization]{A timeline diagram showing the evolution of voice training goals over time. Two parallel horizontal lines represent the progression: the upper orange line shows goal evolution (from 'Descriptive Goal' through 'Initial Technical Goal G0' and subsequent 'Evolving Goals G1, G2'), while the lower green line shows voice progression (from 'Original Voice' through 'Incremental Targets T1, T2, T3' to 'Final Voice'). Vertical orange arrows connect goals to targets, while curved green dashed arrows show progress feedback. The diagram illustrates how end goals inform immediate targets while progress influences goal evolution, creating a dynamic feedback loop in voice training}
    \caption{\added{Voice goals represent the desired trajectory of the voice-training journey, informing how experts and trainees set incremental targets. Although goals evolve beyond the initial goal voice exploration as a result of external factors and skill and knowledge-building with voice training progress, they always envision the end-goal.}} 
    \label{fig:evolving_goals_and_targets}
\end{figure*}

\added{Our research revealed a fundamental gap between how trainees initially conceptualize their voice goals and how experts guide effective voice training. We found that voice goals exist in two distinct forms: \textit{layperson-descriptive goals}, which capture identity presentation needs through layman terminology, and \textit{technical goals}, which provide technical direction for voice training through measurable parameters. This distinction emerged as critical, as descriptive goals, while authentically representing trainees' needs, fail to provide the incremental framework needed for sustainable voice training. In contrast, actionable technical goals enable systematic progress through clear, parameter-based milestones. We found that while trainees and experts work through a process of goal voice exploration as the initial step to voice training, voice goals continue to change along the journey, with skill and knowledge-building playing a key role in evolving trainees' understanding of their goals, as shown in Figure~\ref{fig:evolving_goals_and_targets}.}

\subsection{\added{The Need for Personalized, Actionable Goals}}
\label{need_for_actionable_goals}
\added{Trainees and experts found that voice training technologies and media fell short of providing the necessary training trajectories for voice training by either reducing descriptive goals directly to practice targets or promoting standardized voices that don't address individual needs. We examine these limitations before exploring how experts and trainees bridge the gap between descriptive and actionable goals.}
\subsubsection{\added{Limitations of Technology-Based Goal Setting}}
\label{technology_goals_limitations}
\added{Voice training technologies like Voice Tools \cite{voice_tools} attempt to transform descriptive gender presentation goals directly into pitch-based targets through color-coded ranges (pink for female voice, blue for male voice). While trainees and experts found target-pitch-based technologies like Voice Tools and Voice Pitch Analyzer \cite{voicepitchanalyzer2024} useful for voice exercise guided through \added{responsibly-set incremental} targets \added{and for measuring progress} (T1-T5, T7, T9, T10 and E1, E3, E5, E6),
they noted that the direct mapping of pitch to gender only misleads trainees by overemphasizing pitch (T1-T3, T5, T7-T9 and E1-E3, E5), and creates unsustainable practice targets that conflate goals with daily exercises (T1, T5, T9, T10 and E1, E2, E4, E6). By failing to separate goal-setting from target-setting, these technologies risked reinforcing unhealthy practice patterns and demotivating trainees who struggled to reach far-off goals (T1, T5, T10, and E1, E2, E4, E6).}
\subsubsection{\added{The Problem with Standardized Voices}}

\added{While voice training media provides valuable skill-building resources, its untargeted nature often promotes ``industry-standard'' voices that may not align with trainees' identities (T1, T8-T10). T1 described this ``industry-standard'' voice for transfemme individuals:}

\begin{quote}
\textit{``I'm not going to keep pursuing... the industry standard of `We're going to make you sound like you're AFAB. We're going to make you sound like you're 16'.''}
\end{quote}

\added{Beyond enforcing binary ``passing'' voices, the demographically-limited voices in voice training media provide inauthentic representations of trainees' identities: T9 developed an unintended Californian ``valley girl'' accent from YouTube-based vowel training methods, resulting in others misidentifying her regional background.}

\subsection{\added{Understanding Trainees' Desires}}
\label{understanding_descriptive_goals}

\added{When beginning voice training, trainees express their goals through descriptive language that captures their identity presentation needs. We found that these descriptions typically fall into four categories that inform, but cannot directly guide, voice training: vocal utility, comfort with the voice, surface-level understandings of parameters, and vivid character descriptions.}

\textbf{\textit{Vocal Utility.}}
 \added{Trainees often framed initial goals around how they wanted others to perceive and interact with them.} This ranged from passing as cisgender (T1, T3, T4, T8-T10 and E1, E3, E6) to creating specific social impacts --- T2 desired a voice ``so masculine that it would make his coworkers who misgendered him look silly,'' while T6 sought a voice  \added{that would create some gender ambiguity, being ``gendered as male most of the time'' but occasionally read as female}. For some trainees (T2, T4), having access to a voice that passed as cisgender would allow them to maintain safety in potentially dangerous environments.

\textbf{\textit{Comfort in Voice}}
 \added{The alleviation of voice dysphoria and development of vocal comfort emerged as a critical motivating factor for T1 and T9. Although providing original motivation for some trainees,} experts  \added{noted that} voice dysphoria \added{often functioned as a double-edged sword --- while providing motivation to get away from the current voice, it could also create a}  barrier to \added{practice (E1, E2, E4, E5), especially during voice \textit{generalization},} which involves using the voice-in-training (VIT) in new environments (E1, E4, E5). E1 \added{highlighted} \added{this challenge}:

\begin{quote}
\textit{``[They're] going home to practice, and if what comes out isn't euphoric, it can be ... a barrier to practicing at all.''}
\end{quote}

\textbf{\textit{Parameter-Based Goals}}
Trainees with  \added{some} voice knowledge may begin with goals focused on  specific vocal parameters  (T1, T2, T4, T6, T9 and E4, E6). While  \added{these parameter-based descriptions provide more concrete direction than other descriptive goals,} this kind of  goal \added{often arises} from misconceptions \added{about} the  \added{relative importance} of \added{a} single parameter, \added{particularly} pitch (T5 and T9). Experts \added{worked with these trainees to develop a more holistic} understanding of voice parameters, \added{enabling more nuanced goal-setting} (E1-E3, E4, E6).

\textbf{\textit{Identity Representation}}
Identity representation through voice  \added{provides varying levels of specificity in goal description}.  By definition, gender-affirming voice trainees seek to acquire a voice that is affirming for representing their gender identities (E3-E6). More specifically, some trainees  desire to have a voice that represents their culture or region (T2 and T10).  \added{The most detailed representations often come through character descriptions or voice archetypes that experts can connect to examples} (E1, E2, E5, E6). \added{While some of these descriptions follow voice archetypes, such as the ``pretty boy'' stereotype T10 used to describe his goals, other trainees describe their goals through characters that represent their identity in some way.} \added{E2 recalled a particularly vivid character-based description:}

\begin{quote}
\textit{``I had one client who said that their ideal voice would be someone who could rock a ball gown but then get in a fight in a bodega. It's a very clear character of what she wanted to sound like.''} 
\end{quote}

\added{However, while such character-based descriptions may evoke a clear mental model for trainees and experts, who have extensive experience working with many trainees, these subjective characterizations do not translate directly into concrete vocal parameters or training approaches. Even detailed descriptions like the ``ball gown and bodega'' example, though memorable, do not directly provide actionable information about the specific vocal characteristics the trainee envisions. Through goal voice exploration methods, experts work with trainees to connect these character descriptions to concrete examples and measurable parameters, transforming evocative but abstract descriptions into actionable training goals.}

\input{Figures/translational_barrier}
\subsection{\added{Goal Voice Exploration Methods}}
\label{goal_voice_exploration_methods}

\added{While the descriptive goals provided by trainees could provide vague direction for voice training, experts required more concrete technical goals to} represent an exact direction for voice training (E1-E3, E5, E6)\added{.}  These parameter-based technical goals allow experts to craft individualized training regimens and  \added{develop trainees' understandings of voice} goal characteristics  (E2, E5, E6).  \added{Given the limited technical knowledge of trainees, the variable specificity of layman-descriptive goals, and the need for a trajectory to guide incremental target-setting, experts and trainees must work together collaboratively to develop actionable, parameter-based technical representations of trainees' original goals.}
\added{Our findings revealed three primary strategies that experts and trainees use to bridge the gap between descriptive and actionable goals: subjective satisfaction through voice practice, approximation and role-play of voices with varying qualities to triangulate desired features, and example-driven exploration involving cooperation between the trainee and expert to discover examples that match aspects of the trainees' desired identity presentation.}

\subsubsection{Parameter-focused Exploration}
\added{Parameter-focused exploration emphasizes developing foundational voice skills through guided practice, allowing trainees to discover their goal voice through systematic experimentation and subjective satisfaction with their progress. Experts (E1, E2, E3, E6) guide trainees through exercises focusing on individual voice parameters, helping them understand how each component contributes to overall voice presentation. E6 describes this process of guided exploration:}

\begin{quote}
\textit{``Our voice instrument... has \added{[the]} capacity to do a lot of things, and so I'll usually talk about finding kind of extremes on the map of what the voice can do... Adding resonance in several different ways as well as shifting pitch and understanding that those parameters are separate and they can overlap... I might give out ideas of like character voices to try or placements to try... doing a rating scale of like how close was that to something that feels comfortable for you.''}
\end{quote}

\added{Through these exercises, trainees gain both technical knowledge and practical experience with their vocal instrument. This approach proves particularly valuable for trainees beginning without concrete voice goals, as it helps them discover possibilities while building fundamental skills. E2 highlights the iterative nature of this exploration process:}

\begin{quote}
\textit{``[I] use this analogy of going to a clothing store and trying on different pieces of an outfit and piecing it together. So we're trying on different resonance qualities. We're trying on different pitch centers, different articulation styles... And then kind of piecing together their goal voice.''}
\end{quote}

\added{The systematic nature of parameter-focused exploration prioritizes understanding voice mechanics while promoting continuous reevaluation of voice goals, but fails to define the goal beyond subjective satisfaction with incremental changes. Experts (E1-E3, E5, E6) emphasize that without a clear endpoint, the trial-and-error nature of this method can become time-consuming and potentially demotivating for trainees, particularly during independent practice. Additionally, focusing solely on current physiological capabilities may unnecessarily constrain goal-setting, as voice training can expand these boundaries through consistent practice (E1, E2, E6).}


\subsubsection{Exploration through Approximation and Role-play}
\added{While parameter-focused exploration builds technical understanding through individual components, approximation and role-play provide holistic approaches that integrate multiple voice parameters simultaneously while reducing psychological barriers to practice using useful template examples. Voice actors serve as particularly effective examples in this approach, demonstrating the flexibility of a single vocal instrument across different characterizations (T5, T10 and E6) while also reducing self-consciousness during large voice alterations (E6). For T5, the ability of voice actors to produce new voices
through minor alterations in various vocal parameters (e.g. pitch, resonance, fry) provided her with a starting point
for understanding how these parameters could impact her voice. For testing out new voices during independent practice, E1, E5, and E6 recommended role-playing environments like tabletop gaming sessions, as the playful nature of these environments encourages voice experimentation through ``trying on'' characters.}

\added{Similar to role-play, singing emerged as a powerful tool for independent voice exploration (T2, T9, T10), since it also provides a culturally accepted format for voice experimentation that helps overcome initial psychological barriers (T9, T1). T9 articulates how this approach facilitated her early training:}

\begin{quote}
\textit{``My goal for the first several months was just to not be miserable with it, and so, [for] most of the earlier training, [I] was just singing in the shower.''}
\end{quote}

\added{The accessibility of singing practice extends beyond psychological comfort. T9 mentioned that singing helped her find examples within her current vocal range, allowing for immediate application in training while building confidence. For some trainees, singing also provides culturally-specific voice examples that might be otherwise difficult to access. T2 specifically noted the value of mariachi singers in providing relevant examples for Spanish-language voice training, since he lacked Spanish-speaking examples in his age range.}


\added{While approximation and role-play allow for multi-parameter exploration of voice grounded in the trainee's physiology, this method cannot provide a distant end-goal, as the physiological differences between trainees and voices make exact mimicry difficult (E4). Despite these limitations, the combination of technical practice with reduced psychological barriers makes approximation and role-play valuable components of the goal exploration process, particularly when integrated with other exploratory methods.}

\subsubsection{Example-\added{driven} \added{Exploration}.}
\added{While approximation and role-play leverage examples primarily as tools for understanding voice mechanics, example-driven exploration prioritizes identifying voices that align with trainees' identity goals, regardless of immediate accessibility. This approach focuses first on establishing clear direction and motivation for voice training through examples that resonate with trainees' backgrounds and aspirations, then uses these examples to inform practice strategies.}

\added{Voice examples that inform goals emerge from various aspects of trainees' lives and experiences. Several trainees and experts (T1-T3, T10, E1, E2, E5) identified people from trainees' immediate social circles, such as friends and family members, as valuable reference points. Other trainees found resonance with voices from media and online spaces, including singers, voice actors, and celebrities (T1, T4, T5, T10, E6), as well as both transgender (T5, T9, T10) and cisgender (T10) online personalities.}

\added{These examples serve dual purposes in the training process. Experts (E1-E3) use them as direct points of comparison when working with trainees, while also extracting specific vocal parameters to establish concrete technical targets (all except E4). This technical grounding helps bridge the gap between trainees' descriptive goals and actionable training targets, where examples function as a communication medium between trainee and expert.}

\added{While unburdened by the limitations of the trainee's current physiology and voice training knowledge, examples do not facilitate a sense of ownership over voice, as they are disconnected from the trainee's original voice and physiology. T1 mentioned that taking ownership over their voice training journey meant taking ownership over the limitations of their vocal ``equipment'', while T2, T8, and T9 found ownership through the ability to incrementally shift voice characteristics through practice.}


\subsection{Goals \added{Shape} and Evolve with the Voice Training Journey}
\label{goals_support_and_evolve}
\added{Voice training goals emerge as dynamic constructs that both guide and adapt throughout the gender-affirming voice modification journey. Through analyzing expert evaluations of successful voice training journeys alongside trainees' personal experiences with difficult and evolving goals, we found that effective goal exploration accounted for the dynamic nature of goals through continuous reevaluation and a focus on subjective satisfaction. Furthermore, we identified two key tensions in goal orientation: the challenge of managing sustainable treatment of voice goals during early skill acquisition, and the psychological barriers that emerge when perfectionism intersects with voice generalization.}

\subsubsection{Goals Change.}
While a major focus of goal voice exploration is finding a voice ``end''-goal, voice training goals are hardly ever static. Instead, they dynamically evolve as trainees build new skills, discover new aspects of their identity, and perceive the changing qualities of their voices. Experts and trainees continue reevaluating goals throughout the voice training journey to maintain a goal that represents the trainees' voice needs.

\textbf{\textit{Why Goals Change.}} The suitability of a goal depends on both factors inherent to voice training (e.g. motivation, skill, knowledge, and voice physiology) (T3-T5, T7, T9, and all experts) and other personal factors (e.g. identity and environment). As trainees develop their skills and knowledge, their understanding of what is achievable expands (E4, E5, E6). This awareness leads to a recalibration of goals, as trainees pursue more ambitious or nuanced voice characteristics. The effort a trainee invests in voice training is linked to their motivation, which changes throughout the training journey (T4, T8, T9 and E2, E6). Progress often serves as a powerful motivator, encouraging trainees to set more challenging goals (T9). Conversely, periods of perceived stagnation may lead to a reevaluation of goals, resulting in more realistic or achievable goals (T9, T10 and E1, E2). As trainees develop their identity, trainees shift their goals to align with their evolving sense of self (T1, T4, T7, T10). Changes in a trainee's environment (e.g. work or living situation) can necessitate adjustments to voice goals or require trainees to maintain multiple context-dependent goals (T1, T2, T5-T10 and E1, E2, E5, E6).

\textbf{\textit{Continuous Reevaluation.}} Voice experts emphasized the importance of reevaluating client satisfaction and well-being throughout the goal-setting and training process. Rather than imposing specific targets, most of the experts (all but E4) emphasized the need to check in regularly on clients' personal comfort and satisfaction with their voices. All experts mentioned that they regularly employed subjective satisfaction-measurement techniques, such as surveys at the beginning of each session. This technique not only gave clients agency in exploration of voice goals, but also allowed for dynamic goal-setting that accommodated for skill and knowledge development among clients (E6), increasing range with exercising the vocal muscles (E1, E2, E6), and other factors that changed clients' goals, such as changes in overall transition goals (E2, E5, E6).
This approach allows for flexibility in goal-setting and acknowledges the individual nature of voice goals. Trainees (T1-T4) also recognized the importance of personal satisfaction with the goal as other factors changed through the voice training journey, often adjusting their goals as they progressed. For T2 and T10, the impacts of testosterone allowed them to focus less on goals, instead appreciating the progress and forming their voices around the natural changes. T1 shared that their voice goals changed with their personal relationship with voice:
\begin{quote}
\textit{``[M]aybe that was an achievable goal, but it wasn't really what I wanted anymore. And my goals sort of shifted to having more acceptance with my voice, whatever that means in any form.''}
\end{quote}

\subsubsection{\hazel{Tackling} Rigidity Around Unsustainable \added{Targets}.}
Experts noted that trainees were often set on \added{achieving} unsustainable or unsuitable voice \added{targets}, especially early in the voice training journey (E1-E3, E5, E6). These unsustainable \added{targets reflected a desire to immediately reach the voice end-goal, and} resulted in trainees straining their voices through unhealthy practice (T1, T5, T7, T10 and E1, E3, E5) while also demoralizing trainees \added{by emphasizing} the distance to the \added{end-goal} (T8, T9 and E3). While experts worked with clients to help them understand voice limitations in relation to the \added{end-}goal (E2, E3, E5), this practice could demotivate trainees while also not representing the plasticity of voice. 

As an alternative strategy, experts temporarily de-emphasize the goal voice while focusing on \textbf{incremental targets} (E1-\hazel{E4, E6}). \hazel{Incremental targets break down a goal into concrete and manageable tasks, guiding trainees towards their goal in an accessible manner (E4, E6).} E2 mentioned how they use this technique:
\begin{quote}
    \textit{``I feel like it can start a shame spiral with some people, like `I'm so far away from this goal voice or this person's voice that I was listening to' ... I think in the beginning we might establish, OK, here's where you're starting, [here's where your] fundamental frequency is, and here's where this person that you were listening to is speaking.''}
\end{quote}

\hazel{Trainees often play a significant role in designing their own targets (E1, E6). Many trainees mentioned setting personalized targets of various types, including practicing small phonemes or words (T1, T4, T5, T7-T10), working on a specific voice parameter (T1-3, T5-T10), and focusing on a specific aspect of physiology (T2, T5, T6, T8, T9), such as keeping the larynx elevated when using their voices (T8). Some trainees also set targets related to voice sustainability, gradually increasing the time they were able to use a particular voice (T2, T4, T6, T8-T10). Working towards the goal through incremental targets makes voice training healthier and more achievable (E1, E3, E4, E6), preventing overemphasis of an at-the-time unfeasible end-goal.}

\subsubsection{\hazel{Breaking} the Perfectionist ``Plateau''.}
Voice experts and trainees identified a ``plateau'' to voice training progress following a phase of rapid voice development (T4, T9 and E1, E3). This plateau involved a combination of ``fine-tuning'' work (T4 and E3) and hurdles for voice generalization (E1, E2, E6). As T4 explained, \textit{``Voice training has a really weird learning curve where you start off [and] you don't see a lot of progress and then there's a period where you're like, `oh, I get it now' and then [for] that month you're gonna get a lot of progress and then later on it's a lot of fine-tuning.''} Trainees  struggled with using an imperfect voice-in-training (VIT) outside controlled practice environments when they thought it wouldn't be perfect (T1, T3, T5, T6, T8 and E5, E6)\added{, and fatigue when sustaining the VIT arose as a primary issue mentioned by experts' clients (E1, E2, E4, E6)}. 

To combat \added{apprehension and fatigue associated with} generalization, trainees controlled the contexts where they would use the VIT (T1, T2, T5, T6, T9 and all experts) and all experts recommended starting with practice partners that the trainee felt comfortable with. \hazel{Additionally, both experts and trainees also mentioned actively creating a conducive space for voice training through activities such as role-play or games (E1, E6 and T1). } Furthermore, experts (E1, E4, E5, E6) identified that feedback focused on small achievements could build trainee self-confidence in generalizing their VIT. For example, E5's encouragement focused on specific voice characteristics, \textit{``[G]iving that encouragement that things are sounding good, or you're doing XY and Z well ... [is] setting people up for success knowing that sometimes things are not going to be perfect and that's OK.''}

\subsection{Towards a More Authentic Voice Exploration}
\label{sec:findings:towards_a_goal_voice_discovery_tool}
\added{
Building on 
the participants' prior technology and voice training experiences, we worked with them to envision how technologies could enhance the voice training journey.}

\subsubsection{\added{Grounded Examples through Voice Changers}}
While not always providing natural-sounding voices, basic voice changer technologies --- such as voice filters, desktop applications, and advanced vocal processors --- provided inspiration for voice training through examples grounded in trainees' own voices (T1, T3, T4, T6, T7, T9, T10). We found that the impact of voice changer technologies on goal exploration varied with the effectiveness of the software in producing a natural-sounding voice and the technical knowledge of trainees in navigating difficult interfaces for producing a voice. For trainees with imperfect experiences with voice changers, the produced voice could still provide a direction to work towards (T1, T4, T8), and an understanding of the limitations of shifting single characteristics of voice (T9), or specific characteristics to assimilate into their own voices (T1, T6). \hazel{For example, T9 explained how their perception of the importance of pitch changed after using a pitch-shifting voice changer: 
\begin{quote}
    \textit{``[Shifting pitch with voice changer] leads to inconsistent gendering on on the games that I played... That [experience] played a key part in [making me] realize that pitch is not the end [goal].''} 
\end{quote}}

When voice changers provided adequate voices that allowed trainees to present their identities, trainees felt that the voice changer voice unearthed their own potential, as it demonstrated what voice training could do for them (T1, T10). For T1, positive reactions from others to the modified voice increased internalization, functioning as a motivation for voice training:
\begin{quote}
    \textit{``I did produce something that I think was a very passable cis woman's voice. And I remember sharing that with people and them being like, `I can't believe that was you'... I think I really internalized that people are responding with the sort of shock or disbelief that that could even have been me. And that's the response I want.''}
\end{quote}
\hazel{Conversely, voice changers could create unrealistic expectations for immediate voice training success (T3, T8, T10). T8 frames the ability of voice changers to show what's possible as a trick, since it doesn't represent the actual abilities of the trainee:
\begin{quote}
     \textit{``[Voice changers are] like the filters when you take pictures. You use those filters and think that's really good. But you don't want to fool yourself, I guess. You don't want to encourage yourself to do what you can't.''}
\end{quote}}

\subsubsection{Filtering and Arranging a Voice Database.}
Voice trainees and experts emphasized the need for a large database of voice models, including various gender, cultural, and age characteristics (T3, T6, T7, T9, T10 and E3, E4, E5). Our research team recognized the need for methods to search, arrange, and filter through such a large database. Methods for filtering, arranging and searching through the voice database included (1) measurable-parameter based methods and (2) model tagging systems requiring human intervention. 

\textbf{\textit{Parameter-based Methods.}} Among parameter-based methods, three techniques emerged, including allowing trainees to define ranges for various vocal parameters including pitch, resonance, and intonation (T2-T7, T9), using voice-analysis techniques such as pitch range measurement to automatically determine if a set of goals are feasible and suitable for the trainee (T1, T6-T10), and allowing trainees to edit parameters directly using a parameter-based voice changer (not RVC) to see what sounds good, either working from their current voice or a voice example (T1).

\textbf{\textit{Tagging Methods.}} Among human-description-based tagging methods, desired voice features included voice accent, such as a valley girl or midwestern accent (T3, T7, T8, T9 and E5), voice archetypes, such as an authoritative feminine voice or a ``pretty boy'' voice (T10 and E3, E6), features of voice not yet directly measurable by voice-analysis software, including breathiness, brightness, and scratchiness (T6, T7 and E3, E4, E5), and characteristics of the person who the model is based on, including emotional characteristics, age and gender (T8, T10). To manage tagging of a large database while also giving trainees agency to describe voices using characteristics understandable to themselves, E6 recommended crowdsourcing methods for voice-tagging, such as those employed by the voice training website Sumian Voice \cite{sumian_voice_wiki}.


\subsubsection{Creating Grounded and Unique Voices.} While full-conversion voice changers could provide a database of voice models representing various identities, trainees worried that directly using another person's voice as a goal would create a low sense of ownership (T1, T6), result in a non-unique voice (T7), or create a voice that was too difficult to achieve (T1, T2, T8 and E3-E5). To address this, we explored various voice-changer-inspired techniques for creating unique voices.

\textbf{\textit{Merging Models.}} To address the problem of voice ownership and individualized voice goals, trainees came up with two use-cases for RVC-model merging. The first was merging multiple voice models together from the database to create unique voices that didn't exist before (T5). The second was creating a model of the trainee's current voice and merging it with the desired model to maintain voice characteristics, grounding the goal voice with characteristics from the trainees' original voice (T1).

\textbf{\textit{Simple Voice Changers.}} A voice-parameter-based voice changer could serve as an extra layer on top of an RVC-based voice changer to allow for further customization in creating a unique voice (T1, T8). Also, pulling understandable parameters from the RVC-based voice changer model to feed a more basic parameter-based voice changer, then modifying the current voice using these parameters could create an approximation that maintains characteristics of the original voice (T1).

\textbf{\textit{Visualizing the journey to a goal.}}
To emphasize the goal as a direction rather than an impression to be perfected, we worked with experts and trainees to determine methods for visualizing the journey towards the goal voice model. Among important features mentioned by trainees and experts were (1) voice parameter measurements of the model, such as pitch, resonance, and pitch range (T1, T5, T7, T9, T10 and E1, E4, E5, E6), (2) effort, difficulty, and time approximations representing the current voice's distance from the model (T3, T6), and (3) milestone models as ``steps'' to the model (T3). \hazel{Some trainees mentioned that such journey visualization would allow more personalized target setting (T1, T2), with scaling difficulty as they progress in the training (T8, T9).} To understand the current voice's distance from the model, we discussed (1) merging models between the trainee's current voice and the model voice with varying weights, reducing the current voice weight for increments closer to the model (T3), (2) using voice models that are not directly the end-goal, but currently emulate-able by the trainee for practice in the direction of the goal model (T5, T8), and presenting the journey in a gamified way, such as milestones and rewards used in language-learning applications like Duolingo (T1, T2, T10 and E6).

\hazel{\subsubsection{Tackling Voice Generalization through \added{VR} Environments.}
To tackle the stress and anxiety of using an ``imperfect'' VIT in real life contexts, voice trainees and experts highlighted a wish for a simulated environment for voice training (T1-T10 and E1, E3-E6). Specifically, having a simulated environment would be conducive for conversation practice in social spaces (E4-E6), providing more opportunities for in-depth goal exploration (T6) and new voice practices (E4). Trainees and experts emphasized the importance of a simulated environment in different phases of voice training. At the beginning, when trainees still lack confidence in their VIT, a simulated space could provide a sense of security to the trainees (T6-T8, T9), while towards middle to the end of voice training, a simulated environment could offer more flexibility and opportunities for voice generalization in various contexts (T4, T5). Two trainees and an expert (T6, T8 and E6) specifically mentioned they would use such simulated environment throughout the voice training journey, with E6 underscoring the benefits of using simulated environments as soon as possible:
\begin{quote}
    \textit{``I think it's the sooner the better if the client feels comfortable [using the simulated environment]...  If they have opportunities to test [their voice], and they they feel confident in that system, [that would] be one of the scaffolding kind of opportunities... with those simulated environments.''}
\end{quote}}


%% file: Tables/trainee_technology_use.tex
\begin{table*}[h]
\renewcommand{\arraystretch}{1.3}
\centering
\caption{Technology Use Supporting Voice Training Among Trainee Participants}
\Description[Technology Use for Voice Training Support Among Trainee Participants]{This table documents technology use among ten voice trainees (T1-T10) across three categories: monitoring and practice tools, voice changer experience, and social resources. Most trainees use Voice Tools for monitoring and practice, often supplemented with voice recordings. Voice changer experience varies widely, from basic filters to professional audio engineering background. Social resources predominantly include online platforms, with YouTube channels (particularly TransVoiceLessons), Discord communities, and social media platforms being common sources of voice training media and community support. The data demonstrates diverse technological engagement in voice training, from basic recording tools to online learning communities.}
\label{tab:trainee-tech-usage}
\begin{tabular}{C{0.05\linewidth} C{0.3\linewidth} C{0.25\linewidth} C{0.3\linewidth}}
\toprule
ID & Monitoring and Practice Tools & Voice Changer Experience & Social Resources \\
\midrule
T1 & Voice Tools, Voice recordings, AI assistant (only programmed to respond to VIT) & Digitech voice processor, Audio engineering career background & Discord (Voice Training Media (VTM), lessons, and community), VTM on Instagram Reels, VTM on TikTok, Seattle Voice Lab (VTM), VTM on Youtube \\
\hline
T2 & Pitch grabber, Voice recordings & N/A & Social media resources, VTM on TikTok \\
\hline
T3 & Voice Tools & Basic voice filters (e.g. Snapchat) & Discord (VTM and Community), VTM on YouTube \\
\hline
T4 & Voice Tools (+replay feature) & Pitch-shifting voice changer (unspecified) & Discord and Reddit (VTM and community) \\
\hline
T5 & Voice Tools, Spectrogram & Used unspecified Voice Changer pre-transition for gender identity presentation & Voice training communities on Reddit, TransVoiceLessons on YouTube \\
\hline
T6 & Voice recordings & Basic voice filters & VTM on TikTok \\
\hline
T7 & Voice Tools & AI Voice Changer, Logitech Gaming software (basic audio pre-processing) & VTM on YouTube \\
\hline
T8 & Voice Tools (+replay feature), Spectrogram & Used voice changer (unspecified) with friends for gender presentation online  & TransVoiceLessons on YouTube, Tutorials \\
\hline
T9 & Voice Tools, Voice recordings & VoiceMod & TransVoiceLessons on YouTube, r/transvoice \\
\hline
T10 & Voice Tools, Voice Pitch Analyzer, Karaoke software, Direct audio feedback, Voice Recordings & MorphVOX & Discord and VRChat voice training communities, VTM on Reddit \\
\bottomrule
\end{tabular}
\end{table*}

%% file: Figures/translational_barrier.tex
\begin{figure*}[htbp]
    \centering
    \includegraphics[page=2, width=0.9\textwidth]{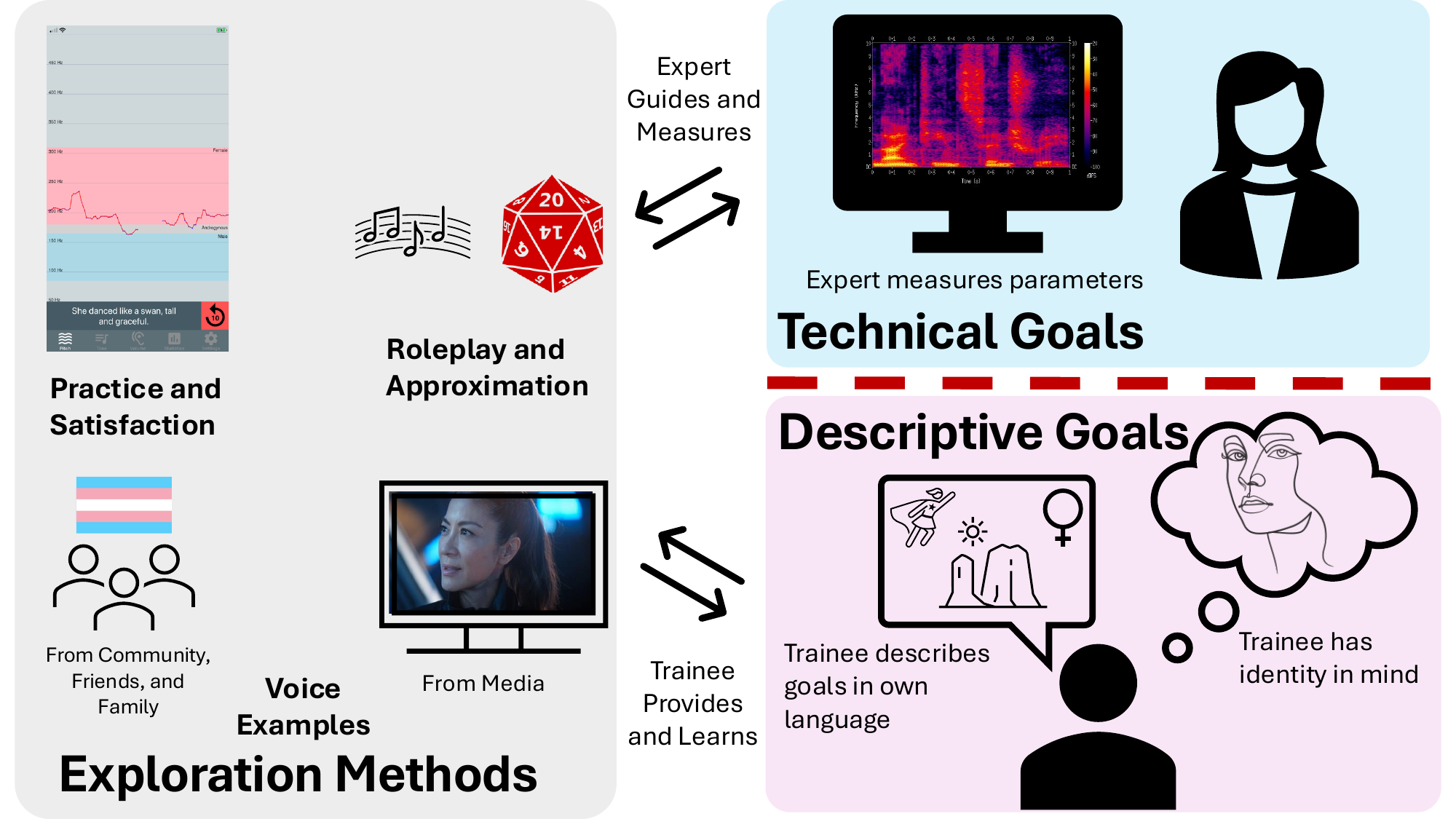}
    \Description[Goal Exploration Methods]{A conceptual diagram showing the voice training process, divided into two main sections: 'Exploration Methods' (left, gray background) and goals (right, blue/pink background). The Exploration Methods section shows three components: 'Practice and Satisfaction' (featuring a waveform graph), 'Roleplay and Approximation' (showing musical notes and a d20 die), and 'Voice Examples' (showing a video screen). These connect via bidirectional arrows to the goals section, which is split into 'Technical Goals' (showing a spectrogram) and 'Descriptive Goals' (showing icons for trainee self-description and identity conceptualization). The layout emphasizes the interactive relationship between exploration methods and goal-setting in voice training whithin the context of cooperative goal exploration between the trainee and expert.}
    \caption{\added{When beginning voice training, trainees start with layperson descriptions of their goals, but defining a clear end-goal trajectory requires technical goals. To define and update the voice-training trajectory, experts and trainees use both subjective-satisfaction with voice production through exercise and measurement and approximation templates (e.g. singers, voice actors, and role-play), and example-driven goal exploration. While subjective satisfaction methods inform the goal as a byproduct of typical voice training methods, example-driven exploration allows experts and trainees to form a vivid understanding of voice goals through the high voice information bandwidth provided through examples chosen based on the trainee's goals.}}
    \Description[Figure showing trainee and expert goal exploration strategies]{This figure }
    \label{translational_barrier}
\end{figure*}

%% file: sections/discussion.tex
\added{Our research provides the first systematic examination of goal voice exploration as a critical component of gender-affirming voice training journeys. Through analysis of expert practices and trainee experiences, we identified three key aspects that characterize successful goal exploration: (1) the dynamic interplay between descriptive goals grounded in identity needs and technical goals enabling concrete practice (Sections~\ref{need_for_actionable_goals},~\ref{understanding_descriptive_goals}), (2) expert and trainee strategies for translating identity presentation desires to actionable goals (Section~\ref{goal_voice_exploration_methods}), (3) the evolution of goals alongside developing knowledge, skills, and identity understanding (Section~\ref{goals_support_and_evolve}), and (4) the potential for novel voice-changer and social VR technologies in aiding the voice training journey as a whole (Section~\ref{sec:findings:towards_a_goal_voice_discovery_tool}).}

\added{Building on these findings, we examine how current technologies' emphasis on standardized targets fails to support the nuanced process of goal exploration revealed in our study. Then, we analyze opportunities to integrate established goal exploration strategies into existing voice training tools and provide examples of potential methods for leveraging novel voice technologies.}
\subsection{\added{Reconceptualizing Voice Goals in Training Technologies}}

\added{
Our findings reveal that successful voice training requires clear separation between long-term goals that provide direction and short-term targets that enable sustainable practice. We suggest three key implications for voice training technology design, including differentiation between target-setting and goal-setting interfaces, approximation of expert strategies for individualized goal exploration, and continuous reevaluation of voice end-goals. 
}

\subsubsection{Emphasizing a Clear Trajectory} We found that the importance of focusing on voice \added{end-}goals changes throughout the voice training journey. Initially, a clear\added{ly defined} goal provides motivation and informs practice indirectly as a concrete end-point. Overemphasis on a \added{technical} goal, such as mistaking \added{technical} goals as voice targets initially in the voice training journey or targeting perfection of the goal, stalls progress. De-emphasizing a goal doesn't remove trajectory from a voice training journey. Instead, it shifts focus to incremental targets and celebration of small achievements, which carry the trajectory. \added{To emphasize this separation while providing personalized guidance, voice training technologies could visualize trainee progress using dynamic ``milestones'' to maintain voice training trajectory and motivation, representing combinations of incremental targets (e.g. sustain, parameter, and generalization targets) with distance informed by trainees' growing skills and expertise. Through these milestones, voice training technologies could incorporate the desired voice training trajectory without over-emphasizing the goal, allowing trainees to craft their journey and track progress sustainably.}


\subsubsection{\added{Individualized} Voice Goals Should Drive Technology-Development}
  Voice training \added{technologies and} media craft voice practice around voice ``standards'' which don't necessarily match the individual needs of trainees. While newer voice training technology research identifies a need for individualized goals \cite{voice_training_software_considerations}, the goal factors outlined \added{do not translate directly to actionable trajectories for voice training, requiring a layer of translation between trainee's descriptive goals and the technical parameter-based goals for setting voice training targets.} We found that \added{translating the descriptive goal of the trainee to an actionable parameter-based goal} is a critical piece of the expert-driven voice training process, allowing experts to tailor voice practice methods to the trainee. Similarly, to provide personalized practice, voice training technologies should triangulate clear voice goals based on the needs of voice trainees. Based on how experts and trainees develop \added{descriptive} goals into \added{actionable} goals, we identify example-based, character triangulation-based, and voice modification methods for goal voice exploration in technologies.

\textbf{\textit{Using Examples.}}
We found that examples provide a potent tool for goal voice exploration early in the voice training process. Without prior knowledge of voice parameters, trainees can use examples to identify desirable voice characteristics.
Similar to goal exploration methods used by experts, a goal exploration system could serve examples and ask trainees to rate examples or identify specific characteristics that they like. Voice parameters such as pitch, pitch variation, and resonance pulled from the examples could inform voice targets, while more specific characteristics of voice such as accent and enunciation could inform individualized voice practice techniques. \added{Furthermore, trainees and experts desired examples that matched their various demographics and diverse gender presentation needs, as opposed to the voice training standards supported by voice training technologies and media. To meet this need, technologies should incorporate large, searchable databases of voice examples that trainees can explore and filter based on descriptive demographic criteria (like gender presentation, age, and cultural characteristics) and experts can understand through pre-measured technical voice parameters.}

\textbf{\textit{Character Triangulation.}}
We found that \added{descriptive} goals capture a range of characteristics that trainees desire for authentic representation, such as vocal utility and cultural identity. When combined, \added{descriptive} goals can \added{help build} a clear\added{er} direction for voice training.
Similar to the expert's guiding role in constructing \added{technical} goals from \added{descriptive demographic-based} goals, a goal exploration system could directly ground goal exploration in a trainee's identity and desired utility for voice. Using a survey of aspects of a trainee's identity, including gender, culture, language, and age, alongside preferred utility, such as passing, gender-nonconformity, or authority, voice training software could identify parameters and practice techniques relevant to developing these vocal identity and utility characteristics. \added{Additionally, when using examples for goal exploration, voice training software should categorize voice examples using both trainee and expert-relevant tags, aiding in the translation of voice archetypes and character descriptions to measurable voice examples. To capture these perspectives, goal exploration technologies could use crowdsourcing from experts and trainees similar to Sumian Voice \cite{sumian_voice_examples}.}

\textbf{\textit{Voice Modification.}}
We found that trainee experiences with basic enhancement-based voice changers, even when imperfect, inform and inspire goal exploration. Voice changers provide a voice example grounded in the trainee's own voice, and, when set up correctly, voice changers provide trainees with feedback for goal exploration based on how others react to the voice. While enhancement-based voice changers have limitations of being difficult to use and producing unnatural and non-passing voices, newer retrieval-based-conversion voice changers \cite{w-okada_voice_changer, RVC_WebUI} have the potential to augment goal exploration through experiences with passing \cite{my_paper}.
Goal exploration systems can use enhancement-based voice changers tied more directly to vocal parameters like pitch, resonance, and weight to develop approximate understandings of vocal parameters into concrete examples of how these parameters would sound when applied to the trainee's voice. Also, through grounding the example in the trainee's current voice at any time during the voice training journey, enhancement-based voice changers could provide a goal reevaluation tool as part of a goal exploration system. Newer retrieval-based conversion systems, while not as grounded, could let trainees effectively test goals as part of the goal exploration process\added{, enable voice-merging for generating unique voices that combine desired features of multiple voices, and partially ground voice examples through merging them with a model trained on the trainee's voice. Finally, by layering enhancement-based voice changers based on percievable voice qualities \cite{permod_robin} on top of RVC voice changers, trainees and experts could fine-tune the goal to create completely new and unique voices that the trainee can form a sense of ownership over.}

\subsubsection{\added{Supporting Trainee Goals through the Nonlinear Voice Training Journey}}  \added{We found that voice training trajectories follow dynamic and individualized paths, as both the end-goal and momentum towards the end-goal change throughout the voice training journey and depend on trainee ability and comfort with using their voice in new settings. To accomodate the individual needs of trainees throughout the journey, voice training technologies should regularly check in with trainees to reevaluate end-goals and incremental targets, while supporting use of the voice-in-training outside of directed practice.} 

\added{\textbf{\textit{Reevaluating satisfaction with the goal.}}} Mirroring expert strategies for goal voice exploration, voice training technologies should incorporate measures of subjective satisfaction, including satisfaction with specific components of the trainee's current voice \added{to continuously reevaluate the voice end-goal. Developers could pick specific milestones informed by inflection points in voice training progress for goal reevaluation --- such as after the trainee builds significant technical knowledge of voice training techniques and their own limits, or when trainees begin using their voice-in-training in new contexts --- or trigger goal reevaluation based on patterns of subjective satisfaction with voice development towards the goal.}

\added{\textbf{\textit{Supporting trainees through the generalization plateau.}} We identified using the voice-in-training in new spaces as a critical challenge for trainees, as the cognitive and social stresses of using the voice in new contexts slow voice training progress significantly. Based on expert strategies, developers should incorporate non-parameter-based progression indicators, providing tools for trainees to track the environments that they use the voice-in-training in, while supporting trainees' voice needs for specific social contexts via context-targeted conversation simulation exercises (e.g., the phrases required for ordering a coffee or talking to a partner). Additionally, developers should consider how their technologies incorporate with online environments which allow trainees to use their voice-in-training safely, with social VR showing potential for allowing trainees to incorporate voice generalization early in their voice training journeys \cite{VR_for_voice_training_and_conversation_simulation, my_paper, freeman_rediscovering_2022}.}

\subsection{Limitations and Future Work}
Although including trainees at various points in their voice training journeys allowed us to  explore a more diverse set of personal voice training experiences, only four trainees had extensive experiences with generalizing their voices across all environments (T2, T3, T4, T9) with two trainees (T1, T8) reporting inconsistent progress with voice training over two years. As a result, while preliminary themes for incremental target-setting and voice generalization arose from expert interviews, not all trainees developed target-setting methods or got to a point where they were generalizing their voice. To better capture target-setting and generalization among trainees, future research should include an additional group of trainees who have fully generalized their ideal voices across all environments. \added{Additionally, while two trainees (T2 and T10) and one expert (E4) provided cultural and ethnic demographic information, as they felt it related to their experiences with voice training and therapy, we did not explicitly ask participants for this information during recruitment or the interview process. Given the unique experiences that these identities created for these participants and the importance of including participants with intersecting identities, future research should include more linguistically, culturally, ethnically, and racially diverse participants to capture their unique voice training journey experiences.}

\added{The key findings of this work focus exclusively on gender-affirming voice training, but the technology implications related to goal discovery could apply beyond this field, such as in identifying desired voices in speech synthesizers for accessibility-related applications or in other speech-language-pathology contexts. Future work should explore how goal voice exploration changes in these contexts.}

Furthermore, while we identify barriers in the voice training journey itself, because of our recruitment criteria limiting trainees to those with more than three months of consistent voice training experience, we did not capture the experiences of transgender individuals with more limited fleeting experiences of voice training. This group could provide information on the early barriers to getting started with voice training that we did not capture through this study. To capture the experiences of this group, future research should include short interviews or surveys focusing on barriers associated with starting voice training.

%% file: sections/conclusion.tex
\added{Through analysis of voice training practices, we identified expert and trainee strategies for discovering, evolving, and using voice goals throughout the training journey. Additionally, we found that end-goals and momentum towards the goal naturally shift alongside trainees' understanding of voice mechanics and identity development, requiring continuous reevaluation to best meet trainee's voice training needs. Building on these insights, we created a set of technology implications and feature examples which show how voice-changer technologies and diverse voice databases can augment traditional goal exploration methods by providing concrete, personalized examples grounded in trainees' voices. This research advances understanding of gender-affirming voice training by exploring how technologies can integrate established voice training practices while leveraging emerging technology capabilities to support successful voice training from setting the initial end-goal to speaking with an authentic voice across social contexts.}

%% file: sections/acknowledgements.tex
This work was partially supported by the National Science Foundation under Grant No. IIS-2328182 and the University of Wisconsin—Madison Office of the Vice Chancellor for Research and Graduate Education with funding from the Wisconsin Alumni Research Foundation.

%% file: sections/Appendix.tex
\section{Appendix}
\include{sections/voice_trainee_interview}
\subsection{Voice Expert Interview Script}
\added{To help answer our research questions, we used the following interview script to conduct semi-structured interviews with experts. Beyond the questions provided here, we regularly asked followup questions for elaboration of close-ended questions and rephrased close-ended questions based on answers to previous questions.}
\label{appendix:expert_interview_script}
\input{sections/voice_expert_interview}
\subsection{Themes and Subthemes}
\input{sections/theme_subtheme_tables}

%% file: sections/voice_trainee_interview.tex
\subsection{Voice Trainee Interview Script}
\label{appendix:trainee_interview_script}
\added{To help answer our research questions, we used the following interview script to conduct semi-structured interviews with trainees. Beyond the questions provided here, we regularly asked followup questions for elaboration of close-ended questions and rephrased close-ended questions based on answers to previous questions.}
\subsubsection{Demographics}
\begin{enumerate}
    \item Do you identify as transgender or gender-nonconforming?
    \item What is your age?
    \item What is your gender identity?
    \begin{enumerate}
        \item Do you identify as transfeminine, transmasculine, or neither?
    \end{enumerate}
    \item How far along are you in your voice training journey (e.g., just starting, midway, almost done)?
    \begin{enumerate}
        \item How long have you been voice training for?
        \item How consistently have you been voice training?
    \end{enumerate}
    \item Have you been working with a voice therapist or voice coach?
    \begin{enumerate}
        \item For how long have you been working with a voice expert?
    \end{enumerate}
\end{enumerate}

\subsubsection{Goal Voice Exploration}
\begin{enumerate}
    \item What was your initial vision, if any, for your voice when you first began voice training?
    \item (If voice goals are simple, such as just ``getting gendered properly'', ask the following question): What models, if any, did you have for what a male/female/other voice would sound like?
    \begin{enumerate}
        \item Did any of these models fit how you wanted to sound?
        \item Why/Why not?
    \end{enumerate}
    \item How did you come to understand or decide on your goal voice?
    \item What voice exploration activities, if any, did you engage in at the start of your voice training to discover a voice goal? (prompts)
    \begin{enumerate}
        \item Pitch glides for determining pitch range
        \item Resonance exercises (e.g. big dog small dog to vocalization)
        \item Mimicking others' and characters' voices
    \end{enumerate}
    \item What factors did you consider when choosing a voice goal? (prompts)
    \begin{enumerate}
        \item Accent of the voice
        \item Perception of the voice's gender
        \item Ease of reaching the voice
        \item How the voice related to your age
        \item How the voice generally fit in with people in your social group
        \item Do you have different goals for different social contexts?
        \item How you'll sound when you're working towards that goal
    \end{enumerate}
    \item How else did you decide on your goal voice?
    \item What was your sense of ownership over your goal voice? (and how has that changed over time?)
    \item Given your current stage of voice training, do you feel that your initial voice goals held up?
    \item If your voice goals changed since you started, how did your goals change?
\end{enumerate}
\subsubsection{Understanding the Goal}
\begin{enumerate}
    \item When you started voice training, what was your understanding of how your goal voice related to your current voice?
    \begin{enumerate}
        \item How did this understanding help or hinder your voice training journey?
        \item How far did you feel you were from your goal voice?
        \begin{enumerate}
            \item How long did you think it would take to reach your voice goals?
            \begin{enumerate}
                \item Do you feel that this is/was accurate?
            \end{enumerate}
            \item How much effort did you think you would have to put into voice training to reach your voice goals?
            \begin{enumerate}
                \item What would you say that the perceived overall difficulty of voice training was?
                \item How frequently did you feel that you needed to voice train?
                \item Do you feel that this was accurate?
            \end{enumerate}
        \end{enumerate}
    \end{enumerate}
    \item Have you set targets for voice training that are not directly the goal, such as incremental targets?
    \begin{enumerate}
        \item What are some examples of these targets that you used? (prompts)
        \begin{enumerate}
            \item Were they focused on single vocal parameters such as pitch or resonance?
            \item Were they focused on keeping up a voice that managed multiple parameters at the same time, such as when reading a short passage?
            \item Were any of them conversational targets?
            \item Were they based on how long you could keep up a trained voice?
            \item Were they based on the distance you could reach from the starting voice (when the target was set?)
            \item In what other ways did you set targets?
        \end{enumerate}
        \item How did you inform these targets?
        \item How achievable were these targets?
        \item How did you integrate these targets into your daily practice?
        \item What would you have improved with voice-target setting?
    \end{enumerate}
\end{enumerate}

\subsubsection{Social Influences on the Goal}
\begin{enumerate}
    \item When setting or refining vocal goals, how did feedback from friends or communities you are part of impact your goals?
    \item When setting or refining vocal goals, how did feedback from voice experts (coaches/therapists) impact your goals?
\end{enumerate}

\subsubsection{Technological Influences on the Goal and Targets}
\begin{enumerate}
    \item How did voice-training related media impact how you set your voice goals?
    \item How did your use of a voice changer impact your voice goals? (prompts)
    \begin{enumerate}
        \item Did you want to sound like the voice changer voice?
        \begin{enumerate}
            \item Why/Why not?
        \end{enumerate}
        \item Did you internalize the voice from the voice changer as your own?
        \item How did your use of a voice changer impact how others associated that voice with you?
    \end{enumerate}
    \item Did you use pitch grabbers (e.g.: voice tools)
    \item How did using tools such as pitch grabbers influence how you set your vocal goals?
    \begin{enumerate}
        \item What information did these tools provide to help you with your vocal goals?
    \end{enumerate}
    \item How did using tools such as pitch grabbers influence how you measured your distance from your goal?
    \item How did using tools such as pitch grabbers influence how you set your vocal targets?
    \begin{enumerate}
        \item How did the information these tools provided help or hinder your development towards vocal targets?
    \end{enumerate}
    \item If voice training support technologies could provide more or clearer representations of information, what would be most useful to you?
    \item If voice training support technologies provided more information, how would that influence your goal and voice-target setting?
    \begin{enumerate}
        \item How would you change how you set your overall goals?
        \item How would you change how you set your incremental targets?
    \end{enumerate}
\end{enumerate}
\subsubsection{Brainstorming}
In this section, we will discuss possible software features for voice training software, and explore the best ways it can meet the needs of voice trainees like yourself. This brainstorming session is divided into three parts, each representing a feature linked to a desired outcome of the software: Goal Voice Discovery, Voice Exercise Towards Incremental Goals, and Simulated Environments for Voice Generalization.

\textbf{\textit{Goal Voice Discovery.}}
Goal Voice Discovery focuses on identifying a goal voice and understanding what the journey to that goal voice will look like. This part of the software makes use of real-time AI voice changers for testing goal voices and feedback based on electronically-detectable vocal characteristics.

\begin{enumerate}
    \item What features of a goal voice would be best for selecting voices from a menu?
    \item When thinking about filters, what characteristics would be most useful?
    \begin{enumerate}
        \item Achievability from the current voice
        \item Fit and feel
        \item Cultural relevancy
        \item Uniqueness
    \end{enumerate}
    \item When addressing whether you can achieve a goal voice, how do you visualize what the journey would be like when working towards that voice?
    \begin{enumerate}
        \item What would aid most in helping visualize what this journey would look like?
    \end{enumerate}
\end{enumerate}

\textbf{\textit{Voice Exercise Towards Incremental Goals.}}
Voice Exercise Towards Incremental Goals focuses on creating healthy short-term voice-training goals and providing the information required to achieve these goals. This part of the software merges your current voice with the goal voice with varying weights to provide a set of vocal parameters at various points.

\begin{enumerate}
    \item When working towards incremental goals, how would you prefer to work towards various parameters?
    \begin{enumerate}
        \item Small jumps towards the goal voice with all parameters (pitch, resonance, vocal weight) worked on at the same time
        \item Large jumps towards the goal voice, with parameters worked on one-at-a-time
        \item A combination of the two --- in what order?
        \item Something else
    \end{enumerate}
    \item Why do you feel that method would work best?
    \begin{enumerate}
        \item Do you have any experiences informing that decision?
    \end{enumerate}
\end{enumerate}

\textbf{\textit{Simulated Environments for Voice Generalization.}}
Simulated Environments for Voice Generalization uses Social VR environments to simulate real-life interactions alongside voice changers as an optional tool for hiding intermediary voice-states. VR overlays provide information on vocal parameters.

\begin{enumerate}
    \item Do you currently use social VR?
    \begin{enumerate}
        \item Do you use social VR as an environment for voice training?
        \item Who do you typically hang around in social VR?
        \begin{enumerate}
            \item Do you voice train or use your trained voice in their presence?
            \begin{enumerate}
                \item Why or why not?
            \end{enumerate}
        \end{enumerate}
    \end{enumerate}
    \item What phases along your voice training would benefit most from having simulated conversation environments?
    \item When voice training, does others hearing your voice feel like a barrier to using your voice-in-training in conversation?
    \begin{enumerate}
        \item (If yes) Would you feel more comfortable if others heard your goal voice from the start of your voice training?
        \begin{enumerate}
            \item Why or why not?
        \end{enumerate}
    \end{enumerate}
    \item What are your worries about using a voice changer in Social VR, if any?
    \begin{enumerate}
        \item How could we address these worries?
    \end{enumerate}
    \item What information would be most critical to helping you maintain a trained voice in a real-life simulation environment if you had a HUD? (prompts)
    \begin{enumerate}
        \item Pitch indicators
        \item Resonance indicators
        \item Basic indicators
        \item Advanced visualizations such as spectrograms
        \item Natural language descriptions
    \end{enumerate}
    \item If you had an overview of how well you hit vocal targets after a session in a conversational environment, what information would be most useful to you?
\end{enumerate}

%% file: sections/voice_expert_interview.tex
\subsubsection{Demographics}
\begin{enumerate}
    \item Do you identify as transgender or gender-nonconforming?
    \item What is your age?
    \item What is your gender identity?
    \item Are you a voice coach or a voice therapist?
    \begin{enumerate}
        \item (If voice therapist) How long have you been practicing?
    \end{enumerate}
    \item How long have you been a voice coach/voice therapist?
\end{enumerate}

\subsubsection{Base Techniques}
\begin{enumerate}
    \item Do clients/students/patients typically have an idea of what their goal voice should sound like when starting voice therapy/training?
    \item Is there an end-goal voice?
    \item In the voice discovery process, what techniques or tools do you typically make use of for helping clients/students/patients find a goal voice?
    \begin{enumerate}
        \item Do you use femininity/masculinity measures?
        \item Do you use model voices from personalities, such as celebrities, characters, or other important figures?
        \item Are there any other ways that you help a client discover a goal voice?
        \item In what ways, if any, might having a goal voice benefit clients/students/patients?
        \item In what ways, if any, might having a goal voice negatively impact clients/students/patients?
    \end{enumerate}
    \item How do you measure the suitability of a goal voice?
    \begin{enumerate}
        \item Do you use femininity/masculinity measures?
        \item Do you measure the ``achievability'' of the goal voice?
        \begin{enumerate}
            \item Do you set intermittent targets if the goal voice is far from the current voice?
            \begin{enumerate}
                \item How do you set these?
            \end{enumerate}
        \end{enumerate}
        \item Do you use goal voice frequently as a measure for success in voice training (that is, comparing current voice vs. goal voice), or do you focus more on other measures? What are those measures?
    \end{enumerate}
    \item What aspects of voice do you typically train for (from highest to lowest importance)?
    \begin{enumerate}
        \item Pitch, resonance?
    \end{enumerate}
    \item What vocal exercise techniques do you typically assign to clients/students/patients for:
    \begin{enumerate}
        \item Pitch control
        \item Resonance control
        \item Other aspects of voice (provide salient examples)
    \end{enumerate}
    \item What would you consider pain-points (specific difficulties) in voice training efforts? Consider:
    \begin{enumerate}
        \item Overall pain-points -- such as problems that cause loss of excitement/energy in voice training
        \item Intermittent pain-points -- topics/exercises that are difficult for clients/students/patients
        \item Anything else that you might have on your mind
    \end{enumerate}
\end{enumerate}

\subsubsection{Technologies Used}
\begin{enumerate}
    \item What are the typical technologies you use during sessions with voice trainees?
    \begin{enumerate}
        \item (Prompts)
        \begin{enumerate}
            \item Pitch analyzers
            \item Voice recording examples from the internet
            \item Voice changers
        \end{enumerate}
        \item What are the uses for each technology?
    \end{enumerate}
    \item Could you share some of the newer or less common technologies you've started using recently?
    \begin{enumerate}
        \item Why do you use these technologies?
        \item (If abandoned) Why did you abandon these technologies?
        \item (If continue using) Have you noticed positive impacts on client/student/patient outcomes when using these technologies?
    \end{enumerate}
    \item Have you recommended students/clients use any technologies when voice training away from your sessions?
    \begin{enumerate}
        \item What sorts of technologies are these?
        \item Do students/clients end up using these technologies?
        \item Have you noticed any problems with these technologies, such as:
        \begin{enumerate}
            \item Unintended negative effects on voice training caused by these technologies (e.g. over-focusing on certain efforts within voice training?)
            \item Student/client difficulties with using or understanding the technology?
        \end{enumerate}
        \item Have you noticed positive impacts in outcomes when patients/clients/students use these technologies?
    \end{enumerate}
\end{enumerate}

\subsubsection{Incorporation of Voice Training into Daily Life}
\begin{enumerate}
    \item What approaches do you recommend for clients to integrate voice training into their daily routines?
    \begin{enumerate}
        \item Could you provide general strategies or tips for clients to use their partially-trained voice in everyday contexts?
    \end{enumerate}
    \item What are common challenges clients face when using their voice-in-training outside of therapy sessions?
    \begin{enumerate}
        \item How do these challenges typically affect their confidence and progress in voice training?
    \end{enumerate}
    \item What does success look like for clients who've effectively integrated voice training into their daily lives?
    \begin{enumerate}
        \item What factors contribute to these successes in your experience?
    \end{enumerate}
    \item What general recommendations do you have for clients to practice their voice in daily conversations?
    \item How critical is regular, real-world practice for the success of voice training?
    \item Are there particular types of activities or environments that are conducive to practicing voice training?
    \item How do you support and encourage clients who may be hesitant to use their partially trained voice in public or social settings?
    \item In general terms, what role do support networks (e.g., friends, family, support groups) play in assisting clients with their voice use outside of therapy sessions?
    \item What methods can clients use to self-monitor and evaluate their voice use in daily life?
    \item Are there tools or techniques you recommend for clients to receive feedback on their voice use outside of therapy sessions?
\end{enumerate}

\subsubsection{Voice Changer Incorporation}
\begin{enumerate}
    \item Have you worked with clients/students/patients who used voice changers for the purpose of changing the perceived gender of their voice online?
    \item Have clients/students/patients incorporated voice changers into their voice training exercises or as a support for using an intermediary (partially-trained) voice?
    \item Do you recommend that anyone uses voice changers as a technology for voice training support?
    \item What's your attitude towards this use of voice changers?
    \begin{enumerate}
        \item Have you considered voice changers to be a supportive, restrictive, or non-relevant technology when it comes to voice training?
    \end{enumerate}
\end{enumerate}

\subsubsection{Brainstorming}
In this section, we will discuss possible software features for voice training software, and explore the best ways it can meet the needs of voice training clients. This brainstorming session is divided into three parts, each representing a feature linked to a desired outcome of the software: Goal Voice Discovery, Voice Exercise Towards Incremental Goals, and Simulated Environments for Voice Generalization.

\textbf{\textit{Goal Voice Discovery.}}
Goal Voice Discovery focuses on identifying a goal voice and understanding what the journey to that goal voice will look like. This part of the software makes use of real-time AI voice changers for testing goal voices and feedback based on electronically-detectable vocal characteristics.

\begin{enumerate}
    \item What features of a goal voice would be best for selecting voices from a menu?
    \item When thinking about filters, what characteristics would be most useful?
    \begin{enumerate}
        \item Achievability from the current voice
        \item Fit and feel
        \item Cultural relevancy
        \item Uniqueness
    \end{enumerate}
    \item When addressing whether a client can achieve a goal voice, how do you visualize what the journey would be like when working towards that voice?
    \begin{enumerate}
        \item What would aid most in helping visualize what this journey would look like?
    \end{enumerate}
\end{enumerate}

\textbf{\textit{Voice Exercise Towards Incremental Goals.}}
Voice Exercise Towards Incremental Goals focuses on creating healthy short-term voice-training goals and providing the information required to achieve these goals. This part of the software merges your current voice with the goal voice with varying weights to provide a set of vocal parameters at various points.

\begin{enumerate}
    \item When working towards incremental goals, how would you prefer clients work towards various parameters?
    \begin{enumerate}
        \item Small jumps towards the goal voice with all parameters (pitch, resonance, vocal weight) worked on at the same time
        \item Large jumps towards the goal voice, with parameters worked on one-at-a-time
        \item A combination of the two
        \item Something else
    \end{enumerate}
    \item Why do you feel that method would work best?
    \begin{enumerate}
        \item Do you have any experiences informing that decision?
    \end{enumerate}
\end{enumerate}

\textbf{\textit{Simulated Environments for Voice Generalization.}}
Simulated Environments for Voice Generalization uses Social VR environments to simulate real-life interactions alongside voice changers as an optional tool for hiding intermediary voice-states. VR overlays provide information on vocal parameters.

\begin{enumerate}
    \item Do you currently use social VR?
    \begin{enumerate}
        \item Do you know any clients who use social VR? Do they use it as an environment for voice training?
    \end{enumerate}
    \item What phases along voice training would benefit most from having simulated conversation environments?
    \item Do clients mention that others hearing their voice feel like a barrier to using their trained voice in conversation?
    \begin{enumerate}
        \item (If yes) Do you think it would be useful if others heard their goal voice from the start of their voice training?
        \begin{enumerate}
            \item Why or why not?
        \end{enumerate}
    \end{enumerate}
    \item What are your worries about clients using a voice changer in Social VR, if any?
    \begin{enumerate}
        \item How could we address these worries?
    \end{enumerate}
    \item What information would be most critical to helping clients maintain a trained voice in a real-life simulation environment if they had a Heads-Up-Display (HUD)? (Prompts)
    \begin{enumerate}
        \item Pitch indicators
        \item Resonance indicators
        \item Basic indicators
        \item Advanced visualizations such as spectrograms
        \item Natural language descriptions
    \end{enumerate}
    \item If you/your client had an overview of how well they hit vocal targets after a session in a conversational environment, what information would be most useful to you/your client?
\end{enumerate}

%% file: sections/theme_subtheme_tables.tex
\renewcommand{\thetable}{A.\arabic{table}}
\setcounter{table}{0}
\begin{table*}[h]
\caption{Voice Training Goals: Themes and Subthemes}
\Description[Table of Themes and Subthemes (Shared and not shared)]{This table presents twelve major themes related to voice training goals, along with their associated subthemes categorized across three groups: shared subthemes (common to both trainees and experts), trainee-specific subthemes, and expert-specific subthemes.

The themes cover technology interactions and practical information about goal exploration:

1. "Voice training technologies fail to provide a personalized training trajectory" highlights shared concerns about pitch-gender mapping and goal conflation, with trainees specifically noting issues with standardization and demographic representation.

2. "Trainees communicating voice desires through their own language" encompasses shared aspects like social interaction needs and character-based identity presentation, with trainees adding cultural and regional influences.

3. "Translating trainee desires into actionable trajectories" features shared methods like role-play and voice actor examples, with experts emphasizing subjective satisfaction alongside exercises.

4. "Grounding the voice in physiology and training experience" addresses equipment limitations (shared) and how personal effort builds ownership (trainee perspective).

5. "Individual goals shaping the voice training journey" focuses on shared aspects of setting incremental targets and deriving personalized targets from end-goals.

6. "Voice end-goals evolving alongside the training journey" shows how both groups reevaluate goals with progress, with trainees noting environmental adaptations and experts highlighting identity development.

7. "De-emphasizing the goal during difficult stages" addresses perfectionism (shared), with experts emphasizing safety and incremental targets.

8. "Voice changers inspiring voice goals" explores how voice changer results shape expectations (shared), with trainees noting both potential and limitations.

9. "Finding a voice that suits my unique identity" combines shared demographic considerations with expert-specific parameter-based search methods.

10. "Fine-tuning a voice to create something unique" focuses on shared aspects like model merging and parameter-based customization.

11. "Mapping the journey through targets" emphasizes shared visualization and gamification approaches, with trainees adding effort estimation perspectives.

12. "Simulated environments enabling easier generalization" primarily addresses shared aspects of conversation practice and scaffolding in safe spaces.

This thematic structure reveals interactions between voice training tools, personal goals, and practical implementation in voice training, while highlighting the shared ideas and different perspectives of trainees and experts.}
\label{tab:voice_training_themes}
\centering
\renewcommand{\arraystretch}{1}
\begin{tabular}{p{0.2\linewidth} p{0.25\linewidth} p{0.25\linewidth} p{0.25\linewidth}}
\toprule
\multicolumn{1}{c}{\large \textbf{Theme}} & 
\multicolumn{1}{c}{\large \textbf{Shared Subthemes}} & 
\multicolumn{1}{c}{\large \textbf{Trainee Subthemes}} & 
\multicolumn{1}{c}{\large \textbf{Expert Subthemes}} \\
\midrule
\added{Voice training technologies fail to provide a personalized training trajectory (The need for a personalized, actionable goal in voice technologies)} & 
\added{Pitch-gender mapping creates misleading focus}\newline
\added{Goals become conflated with daily targets} & 
\added{Industry-standard voices lack personalization}\newline
\added{Training media lacks demographic diversity} &  \\
\hline
\added{Trainees communicating voice desires through their own language (Understanding trainees' desires)} & 
\added{Desired voice utility needs for social interaction}\newline
\added{Voice comfort shapes motivation and barriers}\newline
\added{Parameters frame initial understanding}\newline
\added{Characters describe identity presentation} & 
\added{Culture and region inform voice desires} &  \\
\hline
\added{Translating trainee desires into actionable trajectories (Goal Voice Exploration Methods)} & 
\added{Role-play as freeing and open-ended exploration}\newline
\added{Examples providing a communication medium}\newline
\added{Voice actors as a flexible exploration template}& 
\added{Singing providing an accessible exploration template} & 
\added{Using subjective satisfaction alongside exercise for exploration} \\
\hline
\added{Grounding the voice in my physiology and training experience increases sense of ownership} &
\added{Equipment limitations shape understanding} &
\added{Working for my voice builds ownership} \newline
\\
\hline
\added{Individual goals shaping/personalizing the voice training journey} & 
\added{Setting incremental targets based on individual needs}\newline
\added{Personalized targets emerging from end-goals} \\
\hline
\added{Voice end-goals evolving alongside the voice training journey} & 
\added{Reevaluating the goal with progress} & 
\added{Skills reshape understanding of goals}\newline
\added{Environment necessitates adaptation of goals}\newline
\added{Reevaluating the goal with progress} & 
\added{Identity development shifts desires} \\
\hline
\added{De-emphasizing the goal during difficult stages of the voice training journey} & 
\added{Perfectionism impacts later stages of voice training} & 
&
\added{Emphasizing safety and sustainability in generalization}\newline
\added{Incremental targets take the wheel from end-goals} \\
\hline
\added{Voice changers inspiring voice goals} & 
\added{VC Results shape expectations} & 
\added{VC demonstrates potential}\newline
\added{VC Failures reveal limitations}\newline
\added{VC Feedback motivates progress} & \\
\hline
\added{Finding a voice that suits my unique identity (brainstorming)} & 
\added{Identity characteristics guide filtering}\newline
\added{Demographics ensure representation} &  &
\added{Parameters enable voice search} \newline
\added{Description systems capture nuance} \\
\hline
\added{Fine-tuning a voice to create something unique (brainstorming)} & 
\added{Models merge for uniqueness}\newline
\added{Original characteristics persist}\newline
\added{Parameter-based VC enable customizations} & \\
\hline
\added{Mapping the journey through targets informed by the goal (brainstorming)} & 
\added{Parameters visualizing progress}\newline
\added{Milestones structuring progression}\newline
\added{Gamification maintaining motivation} & 
\added{Effort estimation guiding expectations} & \\
\hline
\added{Simulated environments enabling easier generalization (brainstorming)} & 
\added{Conversation practice in safe, social spaces}\newline
\added{Providing a scaffolding for early practice}\newline &
\added{Providing a flexible environment for generalization}\newline
& \\
\bottomrule
\end{tabular}
\end{table*}